\documentclass[%
reprint,
onecolumn,
frontmatterverbose, 
amsmath,amssymb,amsfonts,
aps,
prx,
floatfix,
noshowkeys,
longbibliography
]{revtex4-2}

\usepackage[T1]{fontenc}

\usepackage{graphicx}

\newcounter{subfigure}[figure]
\renewcommand{\thesubfigure}{\alph{subfigure}}

\newcommand{\sidecaption}[1]
{\rule{0pt}{\abovecaptionskip}
  \refstepcounter{subfigure}%
  \raisebox{-\height}{(\thesubfigure)~}
  \label{#1}\ignorespaces}

\usepackage{mathtools}
\DeclarePairedDelimiter\abs{\lvert}{\rvert}

\usepackage{mleftright} 
\usepackage{braket}
\usepackage{stmaryrd}
\usepackage{siunitx}

\usepackage{float} 
\usepackage{algorithm}
\usepackage{algpseudocodex}

\makeatletter
\def\ftype@algorithm{4}
\makeatother

\algrenewcommand\textproc[1]{\texttt{#1}}

\algnewcommand\algorithmicRepeatTimes{\textbf{repeat}}%

\makeatletter
\algdef{SE}[RepeatTimes]{RepeatTimes}{EndRepeatTimes}[1]{\algpx@startIndent\algpx@startCodeCommand\algorithmicRepeatTimes\ #1\ \algorithmicdo}{\algpx@endIndent\algpx@startCodeCommand\algorithmicend\ \algorithmicRepeatTimes}%

\ifbool{algpx@noEnd}{%
  \algtext*{EndRepeatTimes}%
  \apptocmd{\EndRepeatTimes}{\algpx@endIndent}{}{}%
}{}%

\pretocmd{\RepeatTimes}{\algpx@endCodeCommand}{}{}

\ifbool{algpx@noEnd}{%
  \pretocmd{\EndRepeatTimes}{\algpx@endCodeCommand[1]}{}{}%
}{%
  \pretocmd{\EndRepeatTimes}{\algpx@endCodeCommand[0]}{}{}%
}%
\makeatother

\makeatletter
\newcommand\floatc@apsproduction[2]{%
  \parbox{\hsize}{%
    \centering
    \normalfont\MakeUppercase{#1}.~#2.%
  }\par
}
\newcommand\fs@apsproduction{%
  \def\@fs@cfont{\normalfont}%
  \let\@fs@capt\floatc@apsproduction
  \def\@fs@pre{}%
  \def\@fs@mid{%
    \kern3pt
    \hrule height.4pt depth0pt
    \kern1.2pt
    \hrule height.4pt depth0pt
    \kern4pt
  }%
  \def\@fs@post{%
    \kern4pt
    \hrule height.4pt depth0pt
    \kern1.2pt
    \hrule height.4pt depth0pt
  }%
  \let\@fs@iftopcapt\iftrue
}
\makeatother
\floatstyle{apsproduction}
\restylefloat{algorithm}

\newenvironment{publishedalgorithm}[1][tb]{%
  \begin{figure}[#1]
  \centering
  \begin{minipage}{0.80\linewidth}
  \begin{algorithm}[H]
}{%
  \end{algorithm}
  \end{minipage}
  \end{figure}
}

\usepackage{hyperref}
\hypersetup{hidelinks}
\usepackage{orcidlink}

\newcommand{\secref}[1]{Sec.~\ref{#1}}
\newcommand{\Eqref}[1]{Eq.~\eqref{#1}}

\newcommand{\figref}[1]{Fig.~\ref{#1}}
\newcommand{\algoref}[1]{Algorithm~\ref{#1}}
\newcommand{\lineref}[1]{Line~\ref{#1}}

\DeclareMathOperator*{\argmin}{\arg\min}
\DeclareMathOperator*{\argmax}{\arg\max}
\DeclareMathOperator{\Coset}{\mathcal{C}}
\DeclareMathOperator{\dist}{dist}

\newcommand{\NoIndentState}[1]{\Statex\hspace*{-\algorithmicindent}#1}
\algnewcommand{\Or}{\textbf{or }}
\algnewcommand{\andop}{\textbf{and }}
\algnewcommand{\Not}{\textbf{not }}
\algnewcommand{\Break}{\textbf{break}}

\algnewcommand{\IfThen}[2]{
  \State \algorithmicif\ #1\ \algorithmicthen\ #2}

\graphicspath{{images/}}

\begin{document}

\title{Simulated-annealing decoder for the XZZX code with greedy-matching initialization}

\author{Tatsuya Sakashita\,\orcidlink{0000-0002-3361-2113}}
\email{t-sakashita@phys.s.u-tokyo.ac.jp}
\affiliation{Department of Physics, The University of Tokyo, 7-3-1 Hongo, Bunkyo, Tokyo 113-0033, Japan}


\begin{abstract}
The XZZX code is a variant of the surface code tailored to address biased noise in realistic quantum devices.
We propose a simulated annealing (SA) decoder for the XZZX code.
Our SA decoder is amenable to parallelization because its Markov chain Monte Carlo updates are simple and local.
To initialize SA, we use a recovery configuration produced by our greedy-matching decoder.
Although $Z$-biased noise is commonly assumed in realistic quantum devices, we instead focus on $Y$-biased noise.
Under $Y$-biased noise, the minimum-weight perfect matching (MWPM) decoder becomes suboptimal because it cannot take into account the fact that a $Y$ error contains both $X$ and $Z$ components on the same qubit.
Our numerical simulations for the code capacity noise model, where only data qubits suffer errors, show that our SA decoder achieves higher accuracy than the MWPM decoder.
They also show that, under moderately $Y$-biased noise, our SA decoder achieves accuracy comparable to that of a decoder based on IBM ILOG CPLEX Optimizer (CPLEX), which uses integer programming to find a minimum-energy error configuration consistent with the measured syndrome.
In our greedy-matching decoder, we randomize the tie breaking among equal-weight pairs.
This randomness generates a variety of initial configurations for SA, which improves the convergence of our SA decoder.
These results suggest that combining SA with our greedy-matching initializer is a promising approach to decoding the XZZX code under $Y$-biased noise in the code capacity noise model.
\end{abstract}

\keywords{XZZX code, simulated annealing, biased noise, code capacity noise model, Metropolis algorithm}
\maketitle


\section{Introduction}
\label{sec:intro}

The ultimate goal of quantum computing is to realize fault-tolerant quantum computation with robust error correction mechanisms.
In superconducting qubit platforms, coherence time (i.e., qubit lifetime) is typically on the order of hundreds of microseconds.
The error correction process must be completed within this time frame.
Practical decoders therefore favor simple algorithms to be readily implemented in hardware.

The surface code is one of the most promising topological error correction codes~\cite{kitaev,surface_code_towards}.
It is experimentally feasible because it requires only nearest-neighbor interactions between data and ancilla qubits~\cite{Zhao,Suppressing_experiment,iOlius,google_surface_code,17qubit_experiment}.

In physical systems such as superconducting qubits, optical systems, and ion traps, realistic noise is often biased, meaning that $Z$ (phase-flip) errors occur more frequently than $X$ (bit-flip) errors.
For such biased noise, modifying the stabilizer generators of the surface code leads to alternative codes, such as the tailored surface code (which has only $Y$-type stabilizers)~\cite{tuckett_tailoring} and the XZZX code~\cite{xzzx_nature}.
The XZZX code achieves higher thresholds under $X$-, $Y$-, or $Z$-biased Pauli noise than the conventional surface code.
The conventional surface code is a CSS (Calderbank--Shor--Steane) code~\cite{calderbank_shor,steane} with stabilizer generators consisting solely of $X$-type or $Z$-type operators.
For the XZZX code, there have been proposals for its rotated version~\cite{xzzx_agnostic,rotated_xzzx}, a cluster-state (foliated) construction~\cite{foliated_xzzx}, numerical simulations for optical and superconducting qubits~\cite{Darmawan_xzzx}, and neutral-atom qubits~\cite{neutral_atom}.

The minimum-weight perfect matching (MWPM) decoder~\cite{kitaev} can be executed in polynomial time in the number of syndrome defects and, hence, is widely used for the surface code.
The standard MWPM decoder decomposes decoding into two matching subproblems.
This decomposition does not explicitly capture the correlations between the $X$ and $Z$ components of a $Y$ error, and its performance can therefore degrade when $Y$ errors are non-negligible.

On the other hand, an MCMC (Markov chain Monte Carlo) decoder and a matrix product state (MPS) decoder~\cite{efficient_MLD} exhibit high decoding accuracy by taking into account the probability of $Y$ errors.
For $Z$- and $Y$-biased noise, the MPS decoder has been studied~\cite{ultrahigh}.
However, for a fixed truncation (bond) dimension $\chi$, the MPS decoder scales as $\mathcal{O}(d^2)$ (up to polynomial factors in $\chi$).
In contrast, MCMC-based decoders use local updates with low computational cost, and the computation is easily parallelized.

One of the earliest studies~\cite{Wootton} on MCMC decoders for the CSS surface code employed parallel tempering, while another study~\cite{Hutter} utilized the Metropolis method fixed at the Nishimori inverse temperature.
For an initial error configuration of MCMC, they employed error configurations generated by acting stabilizers and logical operators randomly on the true error configuration~\cite{Wootton, xzzx_agnostic} and a recovery chain obtained by the MWPM decoder~\cite{Hutter}.
Thus, the aim of these previous works was to determine the threshold of the code accurately.

To efficiently minimize energy in MCMC methods, simulated annealing (SA)~\cite{Kirkpatrick} is often used in statistical physics.
By formulating the decoding problem as a random-bond Ising model, SA has been applied to decoders for the CSS surface code~\cite{takeuchi_comparative_decoder} and the color code~\cite{takada}.
Another formulation of an SA decoder for the CSS surface code has a Hamiltonian with penalty terms~\cite{fujii_penalty}, and was demonstrated in a hardware-implemented annealing machine~\cite{fujisaki_DA,fujisaki_circuit-level,takeuchi_comparative_decoder}.

The XZZX code has been studied using the MWPM decoder, the MPS decoder~\cite{xzzx_nature}, and the MCMC-based decoders~\cite{xzzx_agnostic}.
They investigated the decoding accuracy for $Z$-biased noise, which is typical in quantum hardware platforms.
Assuming knowledge of the probabilities of $X,Y,Z$ errors, the authors incorporated this prior information into the weight coefficients of the Hamiltonian so as to improve decoding accuracy.

Reference~\cite{xzzx_agnostic} assumes the code capacity noise model, in which only data qubits suffer independently and identically distributed errors.
To explore configurations efficiently, they carried out the MCMC method at a lower inverse temperature than the Nishimori inverse temperature while keeping the weight coefficients invariant in the Hamiltonian.
Their numerical results suggest that extracting a dominant term of the partition function is sufficient to determine the correct logical operator class.
Thus, their algorithm may be useful for collecting this term, which could then be used as training data for a neural-network decoder.

In this paper, we propose an SA decoder for the XZZX code.
Our goal is not to determine the threshold as accurately as possible, but rather to study whether an SA-based decoder can provide a useful accuracy-run-time trade-off for the XZZX code under $Y$-biased noise.
We assume the code capacity noise model.
Specifically, we focus on $Y$-biased noise, under which the standard MWPM decoder can lose accuracy because its separate treatment of the $X$- and $Z$-error components does not explicitly account for their correlation within a $Y$ error.
In contrast, the SA decoder minimizes an energy function over error configurations, and the relative probabilities of $X$, $Y$, and $Z$ errors are directly incorporated into this energy function.
To initialize SA, we employ a recovery configuration produced by a greedy-matching decoder.
All algorithms are implemented in C++ to minimize decoding time.

Compared with previous MCMC-based studies, the present work differs in three main respects.
First, we formulate an SA decoder for the XZZX code based on a minimum-energy formulation for finding a maximum a posteriori (MAP) error configuration.
Second, we introduce a greedy-matching initializer and a randomized tie-breaking variant to generate diverse initial configurations for independent SA runs.
Third, we focus on $Y$-biased noise, where correlations induced by $Y$ errors make the standard MWPM decoder suboptimal.

The rest of this paper is organized as follows.
In \secref{sec:xzzx_code}, we review the XZZX code.
For the XZZX code, we describe the principles of error detection and correction in \secref{sec:error_detection}, and then present decoding algorithms, which include our SA decoders in \secref{sec:algorithm}.
We show numerical results in \secref{sec:result} and discuss the benefits of our SA decoder in \secref{sec:discussion}.
We conclude in \secref{sec:conclusion}.

\section{The XZZX code}
\label{sec:xzzx_code}

In this section, we briefly review the XZZX surface code~\cite{xzzx_nature}.
We consider a rotated square lattice with open boundaries~\cite{kitaev-open_boundary}.
Data qubits reside on the vertices.

\begin{figure}[tb]
\centering%
\sidecaption{subfig_xzzx:a}\raisebox{-\height}{\includegraphics[scale=0.32]{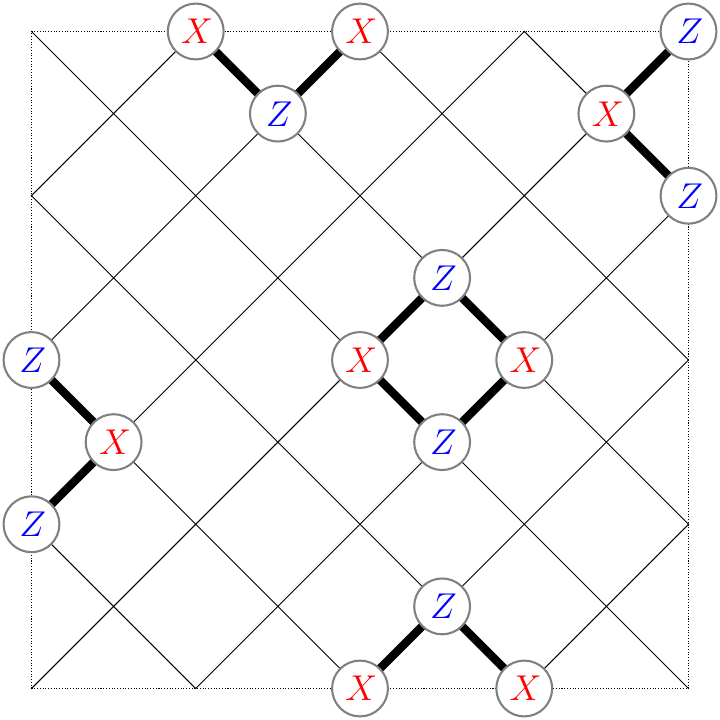}}\\
\sidecaption{subfig_xzzx:b}\raisebox{-\height}{\includegraphics[scale=0.32]{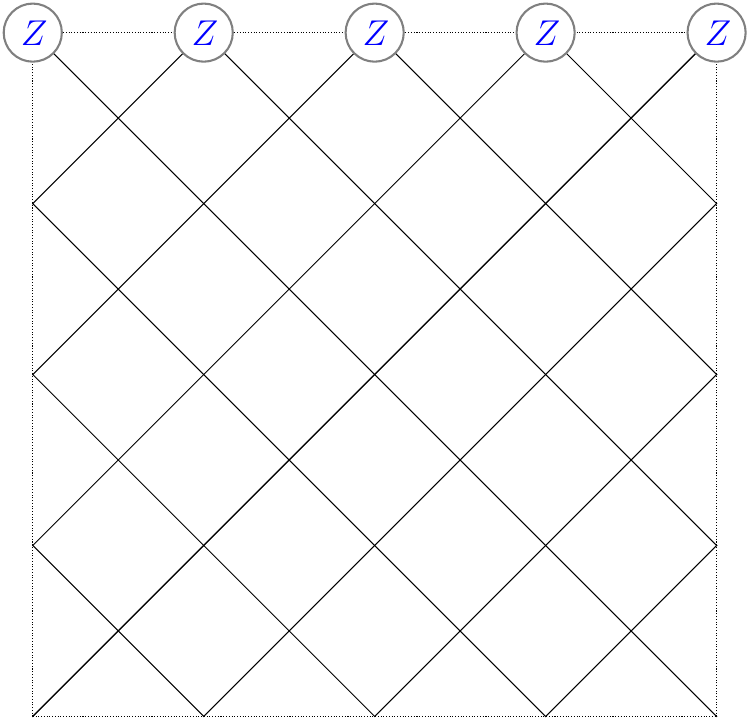}}%
\sidecaption{subfig_xzzx:c}\raisebox{-\height}{\includegraphics[scale=0.32]{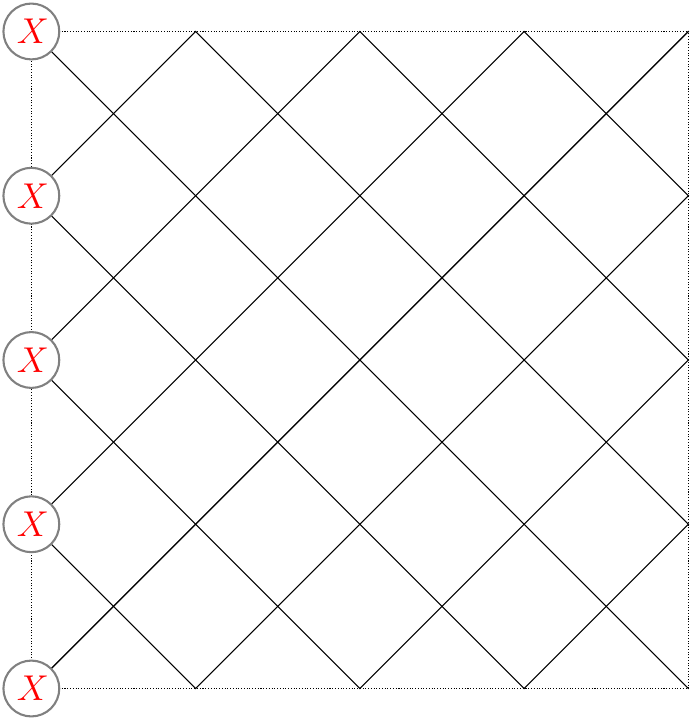}}%
\caption{The XZZX code for the code distance $d=5$: (a) Stabilizer generators $G_f$, showing four-body operators in the bulk and three-body operators at the boundaries. (b) Logical operator $L_Z$. (c) Logical operator $L_X$.\label{fig:XZZX_code}}
\end{figure}
We illustrate stabilizer generators for the XZZX surface code in \figref{fig:XZZX_code}(\ref{subfig_xzzx:a}).
The stabilizer generator acting around a face $f$ is defined as
\begin{equation}
G_f \coloneqq X_{\mathrm{left}(f)} \, Z_{\mathrm{up}(f)} \, Z_{\mathrm{down}(f)} \, X_{\mathrm{right}(f)}
,
\label{eq:stabilizer_generator}
\end{equation}
whose form is the origin of the name ``XZZX code.''
This stabilizer generator is a four-body operator in the bulk and a three-body operator at the boundaries.
The XZZX code has only one type of stabilizer generator, unlike the CSS surface code that has two types of generators.
Since each stabilizer generator includes both $X$ and $Z$ operators as shown in \Eqref{eq:stabilizer_generator}, the XZZX code is not a CSS code.
At the center of every face $f$, an ancilla qubit is placed to measure the stabilizer generator $G_f$ and detect errors.

We define logical operators $L_Z$ ($L_X$) as a Pauli chain connecting the left and right (top and bottom) boundaries, respectively.
Figures~\ref{fig:XZZX_code}(\ref{subfig_xzzx:b}) and \ref{fig:XZZX_code}(\ref{subfig_xzzx:c}) depict examples of the logical operators $L_Z$ and $L_X$, respectively.
We further define $L_Y \coloneqq \mathrm{i} \, L_Z \, L_X$.

Throughout this paper, we ignore global phases of Pauli operators (i.e., overall factors $\pm 1$ and $\pm \mathrm{i}$ are identified).
Let $\overline{\mathcal{P}}$ denote the $N_\mathrm{q}$-qubit Pauli group modulo global phases.
In particular, we identify $L_Y$ with $L_Z L_X$ in $\overline{\mathcal{P}}$.

We identify Pauli chains that differ by multiplication of a stabilizer operator in the stabilizer group $\mathcal{G} \coloneqq \langle G_f \mid f:\text{face} \rangle$.
More precisely, for $P \in \{I,X,Y,Z\}$, we define the logical Pauli operator as the equivalence class $[L_P]$ modulo stabilizers, i.e., $L_P \sim G L_P$ for any $G\in\mathcal{G}$.

The code distance $d$ is defined as the minimum weight of a nontrivial logical operator (equivalently, the length of the shortest Pauli chain representing $L_Z$ or $L_X$).
For the surface codes including the XZZX code, this distance coincides with the linear size of the lattice.
The number of data qubits is $N_\mathrm{q} = d^2 + (d-1)^2$.
The number of stabilizer generators is equal to the number of faces, i.e., $N_\mathrm{stabilizer} = 2d(d-1) = \mathcal{O}(d^2)$.

The XZZX code is advantageous under biased noise.
That is, the XZZX code has a threshold $p_\mathrm{th}=0.5$ for pure $X$, $Y$, or $Z$ noise in the code capacity noise model, reflecting its mapping to a classical repetition code~\cite{xzzx_nature};
on the other hand, the CSS surface code has $p_\mathrm{th}=0.109$ for the pure $X$ or $Z$ noise~\cite{kitaev} and $p_\mathrm{th}=0.5$ for the pure $Y$ noise~\cite{tuckett_tailoring}.

\section{Principles of error detection and correction}
\label{sec:error_detection}

\begin{figure}[tb]
\includegraphics[scale=0.5]{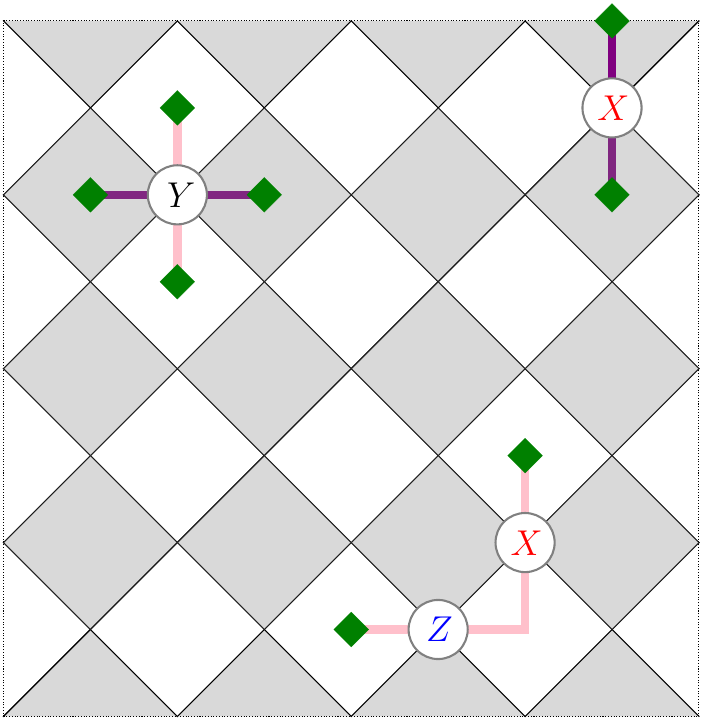}
\caption{An example of the error configuration $C$.
Circles indicate data qubits with $X$, $Y$, or $Z$ errors; qubits without errors are omitted.
The error configuration can be represented as two types of error chains $C_\mathrm{white}$ and $C_\mathrm{gray}$, connecting the white faces and gray faces and drawn with the pink and purple lines on the corresponding decoding graphs $(\mathcal{V}_\mathrm{white}, \mathcal{E}_\mathrm{white})$ and $(\mathcal{V}_\mathrm{gray}, \mathcal{E}_\mathrm{gray})$, respectively.
The end points of these error chains are captured by the syndrome defects, drawn with the green squares.
In both the error chains $C_\mathrm{white}$ and $C_\mathrm{gray}$, the $Z$- and $X$-type Pauli components correspond to operators on data qubits located at the centers of horizontal and vertical edges, respectively.
A $Y$ error occurs precisely on a data qubit where both the $Z$- and $X$-type components are present (i.e., where the corresponding horizontal and vertical edges are both occupied).
\label{fig:XZZX_error_chain_syndrome}}
\end{figure}

Throughout this paper, an \emph{error configuration} $C \in \overline{\mathcal{P}}$ specifies, for each data qubit, whether it experiences no error ($I$) or a Pauli error $X$, $Y$, or $Z$.
Figure~\ref{fig:XZZX_error_chain_syndrome} illustrates an example of the error configuration.
In \figref{fig:XZZX_error_chain_syndrome}, $X,\, Y,\, Z$ errors occur on specific data qubits.

To describe error detection, we introduce two \emph{decoding graphs} $(\mathcal{V}_\mathrm{white}, \mathcal{E}_\mathrm{white})$ and $(\mathcal{V}_\mathrm{gray}, \mathcal{E}_\mathrm{gray})$.
We write $(\mathcal{V}_\bullet, \mathcal{E}_\bullet)$ with $\bullet \in \{\mathrm{white}, \mathrm{gray}\}$ and define it as follows:
\begin{enumerate}
\renewcommand{\labelenumi}{(\arabic{enumi})}
\item \emph{The set of vertices $\mathcal{V}_\bullet$:} The vertices are the centers of faces with the color $\bullet$; each vertex corresponds to an ancilla qubit used to measure the stabilizer.
\item \emph{The set of edges $\mathcal{E}_\bullet$:} The edges connect nearest-neighbor vertices in $\mathcal{V}_\bullet$. Each edge is associated with the data qubit located at its midpoint; an error on that data qubit is represented on the corresponding edge.
\end{enumerate}

The set of edges on which errors occur is called an \emph{error chain}.
There are two types of error chains $C_\mathrm{white} \subset \mathcal{E}_\mathrm{white}$ and $C_\mathrm{gray} \subset \mathcal{E}_\mathrm{gray}$, which connect centers of white and gray faces, respectively, and are drawn with the pink and purple lines in \figref{fig:XZZX_error_chain_syndrome}.
In this paper, to avoid confusion, $C$ (without a subscript) denotes an error configuration, whereas $C_{\bullet}$ (with $\bullet \in \{\mathrm{white}, \mathrm{gray}\}$) denotes an error chain.
For both the error chains $C_\mathrm{white}$ and $C_\mathrm{gray}$, the horizontal and vertical edges represent the presence of $Z$- and $X$-type Pauli components, respectively (i.e., $Z$ or $Y$ on a horizontal edge, and $X$ or $Y$ on a vertical edge).
A $Y$ error occurs exactly at the intersection point of both the error chains $C_\mathrm{white}$ and $C_\mathrm{gray}$.
With a slight abuse of notation, we use the same symbol $C_\bullet$ to denote the corresponding Pauli operator.

For each face $f$, we measure the stabilizer generator $G_f$ and thereby obtain a measurement outcome $s_f \in \{0,1\}$.
For an error chain (viewed as a Pauli operator) $C_\bullet$, the outcome is determined by whether $C_\bullet$ commutes or anticommutes with $G_f$:
\begin{equation}
G_f \, C_\bullet = (-1)^{s_f} \, C_\bullet \, G_f
,
\label{eq:G_f_comm_anticomm}
\end{equation}
i.e., $s_f=0$ if $[G_f,C_\bullet]=0$ and $s_f=1$ if $\{G_f,C_\bullet\}=0$.
We refer to a face $f$ with $s_f = 1$ as a \emph{syndrome defect}.

A syndrome defect appears precisely at an end point of the error chain $C_\bullet$.
More precisely, $s_f = 1$ if and only if $v_f \in \partial C_\bullet$, where $v_f$ denotes the vertex (i.e., face center) corresponding to the face $f$, and $\partial C_\bullet$ denotes the set of vertices in $\mathcal{V}_\bullet$ that have odd incidence with the edge set $C_\bullet$ (equivalently, the end points of the chain $C_\bullet$).

In \figref{fig:XZZX_error_chain_syndrome}, syndrome defects are drawn with green squares.
They capture the end points of $C_\mathrm{white}$ and $C_\mathrm{gray}$.
At such end points of horizontal and vertical edges, $Z$- and $X$-type error components anticommute with $X$ and $Z$ operators acting on the left/right and up/down qubits in the stabilizer generator $G_f$, respectively, and hence can be detected.

For an error chain $C_\bullet$, the set of face centers whose ancilla qubits possess syndrome defects is called the \emph{syndrome} and denoted by
\begin{equation}
S_\bullet \coloneqq \left\{ v_f \in \mathcal{V}_\bullet \, \middle|\, s_f = 1 \right\}
.
\label{eq:def_syndrome}
\end{equation}
The syndrome-consistency condition can then be expressed as
\begin{equation}
\partial{C}_\mathrm{white} = S_\mathrm{white},\qquad  \partial{C}_\mathrm{gray} = S_\mathrm{gray}
.
\label{eq:syndrome_consistency}
\end{equation}

The two error chains $C_\mathrm{white}$ and $C_\mathrm{gray}$ contain the same information as the error configuration $C$.
We express the correspondence as $C = ( C_\mathrm{white}, C_\mathrm{gray} )$ and then define the tuple $\partial{C} \coloneqq ( \partial{C_\mathrm{white}}, \partial{C_\mathrm{gray}} )$.
Similarly, we define the tuple $S \coloneqq ( S_\mathrm{white}, S_\mathrm{gray} )$.
Using this abbreviated notation, the syndrome-consistency condition of \Eqref{eq:syndrome_consistency} reads $\partial{C} = S$.

The true error configuration $\bar{C}$ is unknown;
the only available information is the syndrome $S$, from which decoders infer a \emph{recovery configuration} $C$ such that $\partial{C}=S$.

\section{Decoding algorithms}
\label{sec:algorithm}

We elaborate on our SA decoder in \secref{sec:SA_decoder}.
Subsequently, we describe the MWPM decoder in \secref{sec:MWPM}.
As a simplified version, in \secref{sec:greedy_matching}, we propose a greedy-matching decoder, exploited to construct an initial configuration for the SA decoder.

\subsection{Simulated annealing decoder}
\label{sec:SA_decoder}

We assume \emph{the code capacity noise model} where only data qubits are noisy and syndrome measurements are perfect.
Specifically, each data qubit suffers an error of $X$-, $Y$-, or $Z$-type, independently and identically distributed with the probabilities $p_x$, $p_y$, $p_z$, respectively.
A nonidentity Pauli error occurs with the probability $p \coloneqq p_x + p_y + p_z$, called the \emph{physical error rate}.
The noise acting on each data qubit state $\rho$ is given by the channel
\begin{equation}
\mathcal{E}(\rho) = (1-p) \, \rho \,+\, p_x X\rho X + p_y Y\rho Y + p_z Z\rho Z
.
\label{eq:noise_map}
\end{equation}

For an error configuration $C$, let $n_x,n_y,n_z$ denote the numbers of qubits that suffer $X, Y, Z$ errors, respectively.
The probability of $C$ is then given by
\begin{eqnarray}
\Pr(C) &\coloneqq& {p_x}^{n_x} \, {p_y}^{n_y} \, {p_z}^{n_z} \, (1-p)^{N_\mathrm{q} - (n_x + n_y + n_z)} \notag\\
&=& (1-p)^{N_\mathrm{q}} \,
\left(\frac{p_x}{1-p}\right)^{n_x} \,
\left(\frac{p_y}{1-p}\right)^{n_y} \,
\left(\frac{p_z}{1-p}\right)^{n_z} \notag\\
&=& (1-p)^{N_\mathrm{q}} \, \pi(C) 
,
\label{eq:prob_C}
\end{eqnarray}
where
\begin{eqnarray}
\pi(C)
&\coloneqq&
\left(\frac{p_x}{1-p}\right)^{n_x} \,
\left(\frac{p_y}{1-p}\right)^{n_y} \,
\left(\frac{p_z}{1-p}\right)^{n_z}
\label{eq:pi_C}
\end{eqnarray}
is the (unnormalized) Boltzmann weight of configuration $C$, i.e., $\Pr(C)$ up to the constant factor $(1-p)^{N_\mathrm{q}}$.

For convenience, we reexpress \Eqref{eq:pi_C} using a common base $p/(1-p)$:
\begin{eqnarray}
\pi(C) &=& \left(\frac{p}{1-p}\right)^{\alpha_x n_x  + \alpha_y n_y + \alpha_z n_z}
,
\label{eq:reexpression_prob_C}
\end{eqnarray}
with
\begin{eqnarray}
\alpha_\mu \coloneqq \frac{\ln\!{\left(\frac{p_\mu}{1-p}\right)}}{\ln\!{\left(\frac{p}{1-p}\right)}}
,
\label{eq:coeff_in_energy}
\end{eqnarray}
where $\mu \in \{x, y, z\}$.
We assume $0< p < 1/2$, so that $\ln(p/(1-p)) < 0$.
Knowing or accurately estimating $p_x$, $p_y$, $p_z$ enables minimum-energy decoding for finding a MAP error configuration.

We define the energy of the error configuration $C$ as the weighted sum of the numbers\footnote{%
The numbers $n_x$, $n_y$, and $n_z$ are calculated as follows.
We formulate and implement the Metropolis method with bit representation taking values in the set $\{0, 1\}$, instead of Ising spins taking values in the set $\{1, -1\}$.
In an error configuration, the error on each data qubit is represented by two bits.
The upper and lower bits represent an $X$ and a $Z$ error, respectively.
Thus, $I$, $Z$, $X$, and $Y$ errors are represented as $00$, $01$, $10$, and $11$, respectively.
As such, counting the numbers of qubits that have bit representations $01$, $10$, and $11$ yields $n_z$, $n_x$, and $n_y$, respectively.
}:
\begin{equation}
\mathcal{H}(C) \coloneqq \alpha_x n_x  + \alpha_y n_y + \alpha_z n_z
.
\label{eq:energy}
\end{equation}
Using \Eqref{eq:energy}, \Eqref{eq:reexpression_prob_C} is expressed as
\begin{equation}
\pi(C)
= \left(\frac{p}{1-p}\right)^{\mathcal{H}(C)}
= e^{-\beta_\mathrm{N}\,\mathcal{H}(C)}
,
\label{eq:pi_exp}
\end{equation}
where
\begin{equation}
\beta_\mathrm{N} \coloneqq - \ln\!\left(\frac{p}{1-p}\right)
\qquad
\text{for } 0 < p < 1/2 
.
\label{eq:nishimori_temp}
\end{equation}
We refer to $\beta_\mathrm{N}$ as the Nishimori inverse temperature corresponding to the physical error rate $p$.
Conditioned on the syndrome-consistency $\partial C=S$, we consider the posterior $\Pr(C\mid S)\propto \pi(C)$, which follows from \Eqref{eq:prob_C}.

Given a syndrome $S$, let
\begin{equation}
\Omega_S \coloneqq \left\{ C \in \overline{\mathcal{P}} \,\middle|\, \partial C = S \right\}
\label{eq:Omega_S_def}
\end{equation}
denote the set of syndrome-consistent configurations.
Fix $R(S)\in\Omega_S$, referred to as a \emph{reference configuration}\footnote{Such a fixed $R(S)\in\Omega_S$ is often called a pure error in the literature~\cite{fujii_book}.}.
Then, any error configuration $C\in\Omega_S$ can be decomposed as
\begin{equation}
C = R(S) \, L_P \, G
\label{eq:C_decomposition}
\end{equation}
for some $P\in\{I,X,Y,Z\}$ and some $G \in \mathcal{G}$~\cite{Poulin}.

Left-multiplying both sides of \Eqref{eq:C_decomposition} by $R(S)$ and using $R(S)^2 = I$ (up to phase) gives
\begin{equation}
R(S) \, C = L_P \, G
.
\label{eq:R_C}
\end{equation}

The product $R(S) \, C$ has a trivial syndrome [i.e., $\partial (R(S) \, C) = \emptyset$], so its equivalence class $[R(S) \, C]$ is well defined (as in \secref{sec:xzzx_code}) and equals one of $[L_P]$ with $P\in\{I,X,Y,Z\}$.
We then define, for each $P\in\{I,X,Y,Z\}$, a coset
\begin{equation}
\Coset_{R(S)}(L_P)
\coloneqq
\left\{ C \in \Omega_S \,\middle|\, [R(S)\, C]=[L_P] \right\}.
\label{eq:coset_def}
\end{equation}
Note that $[L_P]$ is an equivalence class in $\overline{\mathcal{P}}/\mathcal{G}$, whereas $\Coset_{R(S)}(L_P)$ is a set of concrete configurations $C \in \Omega_S$;
these objects live in different spaces.
When $S=\emptyset$ and we choose $R(S) = I^{\otimes N_\mathrm{q}}$, for each $P\in\{I,X,Y,Z\}$, $\Coset_{R(S)}(L_P)$ corresponds to the logical operator class $[L_P]$.

The four cosets $\{\Coset_{R(S)}(L_P)\}_{P\in\{I,X,Y,Z\}}$ form a partition of $\Omega_S$:
\begin{equation*}
\Omega_S
=
\Coset_{R(S)}(L_I)\,\dot\cup\,
\Coset_{R(S)}(L_X)\, 
\dot\cup\,
\Coset_{R(S)}(L_Y)\,\dot\cup\,
\Coset_{R(S)}(L_Z),
\end{equation*}
where $\dot\cup$ denotes a disjoint union.
Using Eqs.~\eqref{eq:C_decomposition} and \eqref{eq:coset_def}, we obtain the explicit description
\begin{equation}
\Coset_{R(S)}(L_P)
=
\left\{ R(S)\,L_P\,G \,\middle|\, G\in\mathcal{G} \right\}.
\label{eq:coset_decomp}
\end{equation}

We define the maximum (unnormalized) Boltzmann weight within each coset as
\begin{equation}
\pi^{\max}_{L_P} \coloneqq \max_{C \in \Coset_{R(S)}(L_P)}\pi(C)
.
\label{eq:pi_L_P}
\end{equation}
A configuration $C$ that attains the maximum in \Eqref{eq:pi_L_P} is the most likely configuration in the coset $\Coset_{R(S)}(L_P)$ under the posterior conditioned on $\partial C=S$.
Using \Eqref{eq:pi_exp}, \Eqref{eq:pi_L_P} is expressed as
\begin{equation}
\pi^{\max}_{L_P} = e^{-\beta_\mathrm{N} \, E^{\min}_{L_P}}
,
\label{eq:P_prop_E}
\end{equation}
where
\begin{eqnarray}
E^{\min}_{L_P} &\coloneqq& \min_{C \in \Coset_{R(S)}(L_P)} \mathcal{H}(C) \notag\\
 &=& \min_{G \in \mathcal{G}} \mathcal{H}\mleft( R(S) \, L_P \, G \mright),\label{eq:min_energy}
\end{eqnarray}
which follows from \Eqref{eq:coset_decomp}.

We determine $P^\ast$ by maximizing $\pi^{\max}_{L_P}$ (equivalently, minimizing $E^{\min}_{L_P}$):
\begin{equation}
P^\ast \coloneqq \argmax_{P\in\{I,X,Y,Z\}} \pi^{\max}_{L_P}
= \argmin_{P\in\{I,X,Y,Z\}} E^{\min}_{L_P}
,
\label{eq:determine_P_from_E}
\end{equation}
which follows from \Eqref{eq:P_prop_E}.
This decoding formulation is based on domain-wall constructions in spin-glass models~\cite[Sec.~3.5]{fujii_book}~\cite{hukushima_domain_wall,Chubb_stat_model}.

Ideally, one would perform maximum-likelihood (ML) decoding over logical operator cosets by comparing the coset weights.
We instead use the MAP: We approximate each coset weight $\sum_{C\in\Coset_{R(S)}(L_P)}\pi(C)$ by its largest term  $\pi^{\max}_{L_P}$ defined in \Eqref{eq:P_prop_E}.
This reduces ML decoding over cosets to a minimum-energy search within each coset.

For each $P\in\{I,X,Y,Z\}$, to estimate $E^{\min}_{L_P}$ in \Eqref{eq:min_energy}, we employ SA~\cite{Kirkpatrick}.
For our SA, in the decomposition of \Eqref{eq:C_decomposition}, $C_{L_P} \coloneqq R(S) \, L_P$ is set as an initial configuration for SA, and the stabilizer component $G$ is updated by the Metropolis method to search for the ground state energy $E^{\min}_{L_P}$.

In most of our simulations, we take $R(S)$ to be the recovery configuration returned by our greedy-matching decoder (\secref{sec:greedy_matching}).
Although it has been rigorously proved that SA converges to the optimal solution (under suitable cooling schedules) in the infinite-time limit~\cite{Geman_Geman}, regardless of the initial configuration, in practical finite time, the initial configuration can significantly affect the convergence speed of SA.
We examine this effect in \secref{sec:effect_of_error_configuration}.

We give our SA decoder in \algoref{algo:SA_decoder}.
To estimate $E^{\min}_{L_P}$ effectively, we run simulated annealing multiple times (say, $N_\mathrm{SA}$ times) starting from the identical error configuration $C_{L_P}$ (Lines \ref{line:for_SA} and \ref{line:invoke_SA} of \algoref{algo:SA_decoder}), and select the minimum energy across the runs (\lineref{line:min_among_SA}).
In \lineref{line:return_recovery_chain}, we output the recovery configuration $C_{L_{P^\ast}} = R\, L_{P^\ast}$.%
\footnote{%
We use SA only to estimate $E^{\min}_{L_P}$;
the output recovery configuration is chosen as $R\,L_{P^\ast}$, since any configuration of the inferred logical coset $\Coset_{R(S)}(L_P)$ is equivalent up to stabilizers.
}

Note that, in the nested \algorithmicfor{} loops in Lines \ref{line:for_SA_logical_op} and \ref{line:for_SA} in \algoref{algo:SA_decoder}, each iteration is an independent SA task and, hence, these tasks are trivially parallelizable.
Implementing this parallelization (and quantifying the overhead) is left for future work.

\algoref{algo:SA_decoder} invokes the simulated annealing, shown in \algoref{algo:SA}.
Note that the \texttt{simulated\_annealing} function is called with a copy of the initial configuration.
\algoref{algo:SA} performs the Metropolis method, given in \algoref{algo:metropolis}.

In \lineref{line:metropolis_spin_flip} of \algoref{algo:metropolis}, we update the configuration $C$ by applying a randomly chosen stabilizer generator $G_f$.
\footnote{%
We formulate error configurations by bits, instead of Ising spins.
Hence, multiplying configurations $C\mapsto C \, G_f$ is implemented as bitwise xor (addition modulo two) on the two-bit representation per qubit.
}
This update keeps the syndrome $S$ invariant.

For SA, we introduce a tempered family of unnormalized weights
\begin{equation}
\pi_\beta(C) \coloneqq e^{-\beta\,\mathcal{H}(C)},
\label{eq:pi_beta_def}
\end{equation}
where $\beta > 0$ is treated as a tunable inverse temperature.
In particular, the unnormalized Boltzmann weight corresponding to the physical error rate is recovered at $\beta=\beta_\mathrm{N}$, i.e., $\pi(C) = \pi_{\beta_\mathrm{N}}(C)$, which follows from \Eqref{eq:pi_exp}.
Therefore, $\beta_\mathrm{N}$ provides a natural fixed-temperature choice for MCMC-based decoding~\cite{nishimori_temperature, nishimori_book,Chubb_stat_model}.

In \algoref{algo:SA}, to accelerate convergence of SA, we employ a logarithmic schedule for the inverse temperature, increasing $\beta$ from $\beta_\mathrm{init} \coloneqq 0.9 \, \beta_\mathrm{N}$ to $\beta_\mathrm{final} \coloneqq \beta_\mathrm{N}$:
\begin{equation}
\beta_i \coloneqq \beta_\mathrm{init} \, \left( 1 + \gamma \, \ln i \right)
\quad\text{for $i = 1,\dots,N_\beta$}
,
\label{eq:beta_i}
\end{equation}
with
\begin{equation}
\gamma \coloneqq
\begin{cases}
\displaystyle \left. \left( \dfrac{\beta_\mathrm{final}}{\beta_\mathrm{init}} - 1 \right) \middle/ \ln N_\beta \right. & \text{for } N_\beta \ge 2,\\
0 & \text{for } N_\beta \in \{0,1\}.
\end{cases}
\label{eq:beta_ratio}
\end{equation}
For each inverse temperature $\beta$, we execute the Metropolis step $N_\mathrm{stabilizer}$ times, which is counted as one Monte Carlo sweep, according to the convention.

\begin{publishedalgorithm}[tb]
\caption{Decoder using simulated annealing}
\label{algo:SA_decoder}
\begin{algorithmic}[1]
\Require $S$ : measured syndrome
\Ensure $C$ : recovery configuration, i.e., $\partial{C} = S$
\Function{SA\_decoder}{$S$}
\State $R \gets$ \Call{greedy\_matching}{$S$}
\ForAll {$P \in \{I,X,Y,Z\}$}\label{line:for_SA_logical_op}
   \State $C_{L_P} \coloneqq R \, L_{P}$  \Comment{Initial configuration for SA}  \label{line:adding_logical_op}
   \For {$i = 1$ to $N_\mathrm{SA}$}\label{line:for_SA}
      \State $E^{(i)}_{L_P} \gets$ \Call{simulated\_annealing}{\Call{copy}{$C_{L_P}$}}\label{line:invoke_SA}
   \EndFor
   \State $E_{L_P} \coloneqq \min\limits_{i=1,\dots,N_\mathrm{SA}}{E^{(i)}_{L_P}}$ \Comment{estimate of $E^{\min}_{L_P}$} \label{line:min_among_SA}
\EndFor
\Statex 
\Statex 
\State $P^\ast \coloneqq \argmin\limits_{P \in \{I,X,Y,Z\}}{E_{L_P}}$   \Comment{\Eqref{eq:determine_P_from_E}}
\Statex 
\State \Return $C_{L_{P^\ast}}$  \Comment{defined as $R\,L_{P^\ast}$ in \lineref{line:adding_logical_op}}  \label{line:return_recovery_chain}
\EndFunction
\end{algorithmic}
\end{publishedalgorithm}

\begin{publishedalgorithm}[tb]
\caption{Simulated annealing}
\label{algo:SA}
\begin{algorithmic}[1]
\Require $C$ : error configuration
\Ensure $E_\mathrm{min}$ : an estimate of the minimum energy encountered during annealing
\Function{simulated\_annealing}{$C$}
\State $E \gets \mathcal{H}(C)$  \Comment{\Eqref{eq:energy}}
\State $E_\mathrm{min} \gets E$

\Statex 

\For {$i = 1$ to $N_\beta$}
   \State $\beta \gets \beta_\mathrm{init} \, \left( 1 + \gamma \, \ln{i} \right)$  \Comment{\Eqref{eq:beta_i}}
   \RepeatTimes{$N_\mathrm{stabilizer}$ times} \Comment{one Monte Carlo sweep}  \label{line:for_N_stabilizer}
      \State $\Delta E \gets$ \Call{Metropolis\_step}{$\beta$,\, $C$}
      \State $E \gets E + \Delta E$
      \State $E_\mathrm{min} \gets \min\{E_\mathrm{min},\, E\}$  \Comment{Update min energy}
   \EndRepeatTimes
\EndFor
\Statex 
\State \Return $E_\mathrm{min}$
\EndFunction
\end{algorithmic}
\end{publishedalgorithm}

\begin{publishedalgorithm}[tb]
\caption{Metropolis method that applies a stabilizer generator}
\label{algo:metropolis}
\begin{algorithmic}[1]
\NoIndentState{\textbf{Input:} $\beta$ : inverse temperature}
\NoIndentState{\phantom{\textbf{Input:} }$C$ : error configuration (modified in place)}
\Ensure Energy difference of error configurations
\Function{Metropolis\_step}{$\beta$,\, $C$}
 \State Select a stabilizer generator $G_f$ uniformly at random.
 \State $\Delta E \gets \mathcal{H}(C \, G_f) - \mathcal{H}(C)$
 \Statex 
 \State Draw $r \sim \mathrm{Uniform}(0,1)$.
 \If{$r < \min\{1, \exp(-\beta \, \Delta E)\}$}
  \State $C \gets C \, G_f$  \Comment{Applies $G_f$ (by bit flips)}  \label{line:metropolis_spin_flip}
  \State \Return $\Delta E$
 \Else \State \Return $0$
 \EndIf
\EndFunction
\end{algorithmic}
\end{publishedalgorithm}

\subsection{The MWPM decoder}
\label{sec:MWPM}

\subsubsection{The MWPM decoding algorithm}
\label{subsec:MWPM_algorithm}

We briefly review the algorithm of the MWPM decoder for the XZZX code~\cite{xzzx_nature}.

The MWPM decoder is formulated on the decoding graph $(\mathcal{V}_\bullet, \mathcal{E}_\bullet)$, introduced in \secref{sec:error_detection}.
We index the vertices by integer coordinates as
\begin{eqnarray*}
\mathcal{V}_\mathrm{white} &= \{(i,j)\mid i\in\{1,\dots,d\},\; j\in\{1,\dots,d-1\}\},\\
\mathcal{V}_\mathrm{gray} &= \{(i,j)\mid i\in\{1,\dots,d-1\},\; j\in\{1,\dots,d\}\},
\end{eqnarray*}
where $i$ and $j$ denote the horizontal and vertical coordinates, respectively, and $(1,1)$ is the lower-left corner.

For vertices $u,v \in S_\bullet$, we define the weight $w(u,v)$ between $u=(u_x,u_y)$ and $v=(v_x,v_y)$:
\begin{equation}
w(u,v) \coloneqq
\min\left\{
w_\mathrm{int}(u,v),\,
w_\mathrm{bnd}(u,v)
\right\}
.
\label{eq:distance_u_v_def_new}
\end{equation}
Here, $w_\mathrm{int}(u,v)$ is the weighted Manhattan distance~\cite{xzzx_nature}
\begin{equation}
w_\mathrm{int}(u,v) \coloneqq \tilde\alpha_z \abs{u_x - v_x} \,+\, \tilde\alpha_x \abs{u_y - v_y}
,
\label{eq:distance_u_v_internal}
\end{equation}
with
\begin{eqnarray}
\tilde\alpha_\mu \coloneqq -\ln\!{\left(\frac{\tilde{p}_\mu}{1-p}\right)}
\quad
\text{for $\mu \in \{x,z\}$}
,
\label{eq:coeff_in_distance}
\end{eqnarray}
where we defined the probabilities $\tilde{p}_x \coloneqq p_x+p_y$ and $\tilde{p}_z \coloneqq p_z+p_y$.
Note that MWPM is invariant under multiplying all edge weights by a positive constant.
Therefore, we may freely rescale the weights without changing the matching solution.

The alternative weight $w_\mathrm{bnd}(u,v)$ is the weight of connecting both vertices to the nearest boundary~\cite{MWPM_complexity}:
\begin{enumerate}
\renewcommand{\labelenumi}{(\arabic{enumi})}
\item For $u,v \in S_\mathrm{white}$, we define
\begin{equation}
\begin{split}
w_\mathrm{bnd}(u,v) \coloneqq
\tilde\alpha_x (
&\min\{\dist_\mathrm{B}(u), \dist_\mathrm{T}(u)\} \, + \\
&\min\{\dist_\mathrm{B}(v), \dist_\mathrm{T}(v)\} \, )
,
\end{split}
\label{eq:distance_u_v_virtual_white}
\end{equation}
where $\dist_\mathrm{B}(u) \coloneqq u_y$ and $\dist_\mathrm{T}(u) \coloneqq d - u_y$ are the vertical distances from the vertex $u$ to the bottom and top boundaries, respectively, and similarly for $v$.
\item For $u,v\in S_{\mathrm{gray}}$, we define
\begin{equation}
\begin{split}
w_\mathrm{bnd}(u,v) \coloneqq
\tilde\alpha_z (
&\min\{\dist_\mathrm{L}(u), \dist_\mathrm{R}(u)\} \, + \\
&\min\{\dist_\mathrm{L}(v), \dist_\mathrm{R}(v)\} \, )
,
\end{split}
\label{eq:distance_u_v_virtual_gray}
\end{equation}
where $\dist_\mathrm{L}(u) \coloneqq u_x$ and $\dist_\mathrm{R}(u) \coloneqq d - u_x$ are the horizontal distances from the vertex $u$ to the left and right boundaries, respectively, and similarly for $v$.
\end{enumerate}

We present the algorithm of the MWPM decoder in \algoref{algo:MWPM_decoder}.
In \lineref{line:if_S_odd}, if the number of vertices $|S_\bullet|$ is odd, a \emph{virtual boundary vertex} $b$ is appended to the syndrome $S_\bullet$ so that a perfect matching exists (i.e., $|S_\bullet|$ becomes even)~\cite{MWPM_complexity}.
In this case, we extend the weight function $w(\cdot,\cdot)$ by defining, for $u\in S_\bullet$,
\begin{equation}
w(u,b) \coloneqq
\begin{cases}
\tilde\alpha_x \min\{\dist_{\mathrm{B}}(u), \dist_{\mathrm{T}}(u)\}, & \text{for } \bullet=\mathrm{white},\\
\tilde\alpha_z \min\{\dist_{\mathrm{L}}(u), \dist_{\mathrm{R}}(u)\}, & \text{for } \bullet=\mathrm{gray}.
\end{cases}
\label{eq:distance_u_boundary}
\end{equation}
Note that $w_\mathrm{bnd}(u,v) = w(u,b) + w(v,b)$, corresponding to pairing both the defects $u,v$ with the boundary $b$.

In \lineref{line:create_K}, we create the set of (unordered) pairs
\begin{equation}
K \coloneqq \{ \{u,v\}  \,\mid\,  u,v \in S_\bullet,\, u \neq v \}
,
\label{eq:list_K}
\end{equation}
which are all possible candidates for pairing.
In \lineref{line:MWPM_graph}, the \texttt{MWPM\_graph} function finds an MWPM~\cite{Edmonds}, i.e., a perfect matching $L \subset K$ such that the total distance $\sum_{\{u,v\} \in L} w(u,v)$ is minimum.
For our implementation, we employ the C++ library, Blossom V~\cite{BlossomV}.

\begin{publishedalgorithm}[tb]
\caption{The MWPM decoder}
\label{algo:MWPM_decoder}
\begin{algorithmic}[1]
\Require $S_\bullet$ : Measured syndrome with $\bullet \in \{\mathrm{white}, \mathrm{gray}\}$
\Ensure $C_\bullet$ : Recovery chain, i.e., $\partial{C_\bullet} = S_\bullet$
\Function{MWPM\_chain}{$S_\bullet$}

\If{$|S_\bullet|$ is odd}  \label{line:if_S_odd}
  \State Append the virtual boundary vertex $b$ to $S_\bullet$
\EndIf

\State $K \coloneqq \{ \{u,v\} \,\mid\,  u,v \in S_\bullet,\, u \neq v \}$  \Comment{\Eqref{eq:list_K}}  \label{line:create_K}
\Statex 

\State $L \gets$  \Call{MWPM\_graph}{$K$}  \Comment{Utilize Blossom V}  \label{line:MWPM_graph}
\Statex 

\State $C_\bullet \gets \emptyset$
\ForAll{$\{u,v\} \in L$}
  \State Add to $C_\bullet$ the edges along a path connecting $u$ and $v$ with the weight $w(u,v)$.\label{line:MWPM_connect}
\EndFor
\Statex 
\State \Return $C_\bullet$
\EndFunction
\end{algorithmic}

\vspace{\baselineskip}

\begin{algorithmic}[1]
\Require $S \coloneqq (S_\mathrm{white},\, S_\mathrm{gray})$ : Measured syndrome
\Ensure $C$ : Recovery configuration, i.e., $\partial{C} = S$
\Function{MWPM}{$S$}
\State $C_\mathrm{white} \gets$ \Call{MWPM\_chain}{$S_\mathrm{white}$}
\State $C_\mathrm{gray} \gets$ \Call{MWPM\_chain}{$S_\mathrm{gray}$}
\State $C \coloneqq (C_\mathrm{white}, C_\mathrm{gray})$
\Statex 
\State \Return $C$
\EndFunction
\end{algorithmic}
\end{publishedalgorithm}

In \lineref{line:MWPM_connect}, for each matched pair $\{u,v\} \in L$, we connect the vertices $u$ and $v$ with a path that consists of edges, i.e., applying $Z$ and $X$ operators to the qubits located at the centers of horizontal and vertical edges, respectively.
This path has weight $w(u,v)$, i.e., it is chosen to attain the distance defined by Eqs.~\eqref{eq:distance_u_v_internal} and \eqref{eq:distance_u_v_virtual_white}--\eqref{eq:distance_u_boundary}.
The set of these edges forms a recovery chain $C_\bullet$.

The \texttt{MWPM} function invokes the \texttt{MWPM\_chain} function twice, i.e., for the syndromes $S_\mathrm{white}$ and $S_\mathrm{gray}$ on the white and gray faces, respectively.
Note that only syndromes on faces of the same color may be connected.

Implementing the MWPM decoder efficiently on hardware such as field-programmable gate arrays (FPGAs) is challenging due to its algorithmic complexity and memory/communication requirements.
Thus, we propose a simple variant algorithm in \secref{sec:greedy_matching}.

\subsubsection{The MWPM decoder versus our SA decoder as a minimization problem}
\label{subsec:MWPM_as_minimization}

Given a perfect matching $L \subset K$, we define its total weight as
\begin{equation}
w(L) \coloneqq \sum_{\{u,v\} \in L} w(u,v)
.
\label{eq:distance_L_MWPM}
\end{equation}
The MWPM decoder finds a minimum-weight perfect matching $L^\ast$:
\begin{equation*}
L^\ast \in \arg\min_{L}\left\{ w(L) \,\middle|\, L \text{ is a perfect matching on } S_\bullet \right\}
.
\end{equation*}

From a perfect matching $L$, we can construct a recovery chain $C_\bullet(L)$, i.e., $\partial{C_\bullet(L)} = S_\bullet$, by connecting each matched pair $\{u,v\}\in L$ with a path of weight $w(u,v)$\footnote{Note that $C_\bullet(L)$ is not unique in general, due to degeneracies in the minimum-weight paths.}.
Since the weighted distance $w(u,v)$ is defined by Eqs.~\eqref{eq:distance_u_v_internal} and \eqref{eq:distance_u_v_virtual_white}--\eqref{eq:distance_u_boundary}, \Eqref{eq:distance_L_MWPM} is expressed as
\begin{equation}
w(L) = \tilde\alpha_z \, n_{\mathrm{hor}} \, + \,
\tilde\alpha_x \, n_{\mathrm{ver}}
,
\label{eq:weighted_distance_L}
\end{equation}
where $n_\mathrm{hor}$ and $n_\mathrm{ver}$ are the numbers of occupied horizontal and vertical edges in the error chain $C_\bullet(L)$, i.e., the numbers of $Z$- and $X$-type components, respectively.
A $Y$ error contributes to both $n_\mathrm{hor}$ and $n_\mathrm{ver}$.

We now compare the total distance in \Eqref{eq:weighted_distance_L} and the energy defined in \Eqref{eq:energy}.
In the MWPM approach for the XZZX code, one solves two separate matching problems on the white and gray decoding graphs.
This effectively decouples the inference of the $X$- and $Z$-type components and does not model the fact that a $Y$ error simultaneously contributes to both components on the same data qubit;
hence it cannot capture $Y$-induced correlations.
Consequently, when $p_y$ is non-negligible, this MWPM approach is expected to be suboptimal compared to our SA decoder, which explicitly incorporates $p_y$ via the $Y$-error weight (through $\alpha_y$) in the energy $\mathcal{H}(C)$ in \Eqref{eq:energy}.

\subsection{The greedy-matching decoder}
\label{sec:greedy_matching}

As described in \secref{subsec:MWPM_algorithm}, the MWPM decoder utilizes the MWPM graph algorithm, i.e., the perfect matching giving the minimum of the total distance.
By replacing the global minimization of MWPM with a local greedy-matching rule, we obtain a simple and low computational cost decoder.
We employ the greedy-matching decoding algorithm to construct a reference configuration $R(S)$, set as an initial configuration for our SA decoder in \algoref{algo:SA_decoder}.

We give our greedy-matching decoding algorithm in \algoref{algo:greedy_matching}.
In \lineref{line:pick_minimum}, the \texttt{greedy\_matching\_chain} function matches the closest pair $\{u,v\} \in K$ with respect to $w(u,v)$, and connects the vertices $u$ and $v$.
\lineref{line:remove_pairs} removes all pairs that contain $u$ or $v$ from the candidate list $K$ of vertex pairs.
This algorithm thus enables us to monotonically reduce the size of the list $K$.

We efficiently implement $K$ using a \emph{priority queue}~\cite[Sec.~6.5]{algorithm_Cormen}.
In the priority queue with $m$ elements (e.g., a binary heap), each of insertion and extract-min operation takes $\mathcal{O}(\log m)$; therefore, performing the operations for $m$ elements has time complexity $\mathcal{O}(m\log m)$.
For our greedy-matching decoder, the size of the priority queue is $m=\mathcal{O}(|S_\bullet|^2)$, where $|S_\bullet|$ denotes the number of syndrome defects.
Thus, the worst-case time complexity is $\mathcal{O}(|S_\bullet|^2\log |S_\bullet|)$ for each color.
Since $|S_\bullet|\le |\mathcal{V}_\bullet|=\mathcal{O}(d^2)$, this yields the worst-case bound $\mathcal{O}(d^4\log d)$.

\begin{publishedalgorithm}[tb]
\caption{The greedy-matching decoder}
\label{algo:greedy_matching}
\begin{algorithmic}[1]
\Require $S_\bullet$ : Measured syndrome with $\bullet \in \{\mathrm{white}, \mathrm{gray}\}$
\Ensure $C_\bullet$ : Recovery chain, i.e., $\partial{C_\bullet} = S_\bullet$
\Function{greedy\_matching\_chain}{$S_\bullet$}
\If{$|S_\bullet|$ is odd}
  \State Append the virtual boundary vertex $b$ to $S_\bullet$
\EndIf

\State $K \coloneqq \{ \{u,v\} \,|\,  u,v \in S_\bullet,\, u \neq v\}$  \Comment{\Eqref{eq:list_K}}
\State $C_\bullet \gets \emptyset$
\Statex 
\While{$K$ is not empty}
  \State Pick a pair $\{u,v\} \in K$ such that $w(u,v)$ is minimum.\label{line:pick_minimum}
  \State Add to $C_\bullet$ the edges along a path connecting $u$ and $v$ with the weight $w(u,v)$.
  \State Remove from $K$ all pairs $\{x,y\}$ such that $x\in\{u,v\}$ or $y\in\{u,v\}$.\label{line:remove_pairs}
\EndWhile
\Statex 
\State \Return $C_\bullet$
\EndFunction
\end{algorithmic}

\vspace{\baselineskip}

\begin{algorithmic}[1]
\Require $S \coloneqq (S_\mathrm{white},\, S_\mathrm{gray})$ : Measured syndrome
\Ensure $C$ : Recovery configuration, i.e., $\partial{C} = S$
\Function{greedy\_matching}{$S$}
\State $C_\mathrm{white} \gets$ \Call{greedy\_matching\_chain}{$S_\mathrm{white}$}
\State $C_\mathrm{gray} \gets$ \Call{greedy\_matching\_chain}{$S_\mathrm{gray}$}
\State $C \coloneqq (C_\mathrm{white}, C_\mathrm{gray})$
\Statex 
\State \Return $C$
\EndFunction
\end{algorithmic}
\end{publishedalgorithm}

Like the MWPM decoder, our greedy-matching decoder employs the weighted distance $w(u,v)$, which is calculated from $p_z$ and $p_x$, and therefore cannot take $p_y$ into account.

\subsection{Our SA decoder that starts with different reference configurations}
\label{sec:SA_decoder_greedy_matching_with_different_distance}

In \secref{sec:SA_decoder}, we described our SA decoder in \algoref{algo:SA_decoder}, which performs SA multiple times to achieve energy as low as possible.

To improve efficiency, rather than using an identical reference configuration repeatedly, it is preferable to use a different reference configuration for each SA run.
To this end, we modify our greedy-matching algorithm by randomizing the tie breaking among equal-weight pairs.
Specifically, we randomize tie breaking among equal-weight pairs using pseudorandom numbers.
This modified algorithm is presented in \algoref{algo:greedy_matching_different}.
In \algoref{algo:greedy_matching_different}, we form the subset
\begin{equation}
K_{\min} \coloneqq \{ \{u,v\} \in K \,\mid\, w(u,v) = w_{\min} \}
,
\label{eq:list_K_min}
\end{equation}
i.e., the set of candidate pairs that attain the current minimum weight $w_{\min}$.

This newly introduced randomness may produce a different reference configuration for every invocation, while still satisfying $\partial{C} = S$.
However, two such reference configurations can differ by both a stabilizer component $G$ and a logical operator component $L_P$ in the decomposition of \Eqref{eq:C_decomposition}.
Therefore, to compare energies across runs consistently, we relabel the logical operator cosets obtained in each run as those defined with respect to a fixed ``standard'' reference configuration.

We present a second SA decoder that utilizes \algoref{algo:greedy_matching_different} in \algoref{algo:SA_decoder_different_error_configurations}.
In \lineref{line:standard_error_configuration}, we construct a reference configuration $R_1(S) \in \Omega_S$, which we use as the fixed standard reference.

Let $R_i(S) \in \Omega_S$ ($i\ge2$) denote the reference configuration produced in the $i$th run.
Using \Eqref{eq:C_decomposition}, $R_i(S)$ can be decomposed as
\begin{equation}
R_i(S) = R_1(S) \, L_Q \, G'
\label{eq:R_i_decomposition}
\end{equation}
for some $Q \in \{I,X,Y,Z\}$ and some $G' \in \mathcal{G}$. 
Since $\partial(R_i(S) \, R_1(S)) = \emptyset$, the equivalence class $[R_i(S) \, R_1(S)] \in \overline{\mathcal{P}}/\mathcal{G}$ is well defined and we obtain $[R_i(S) \, R_1(S)] = [L_Q]$.
The Pauli operator $Q$ can be determined by checking its commutation relations with the logical operators $L_X$ and $L_Z$.

Since stabilizers commute with logical operators, \Eqref{eq:R_i_decomposition} implies the correspondence
\begin{equation}
\Coset_{R_i(S)}(L_P) = \Coset_{R_1(S)}(L_{Q P})
,
\label{eq:coset_correspondence}
\end{equation}
where $QP$ denotes the Pauli multiplication in $\{I,X,Y,Z\}$ (up to global phase), consistent with our convention of ignoring global phases.
Equation~\eqref{eq:coset_correspondence} means that ``the coset of $L_P$ with respect to $R_i(S)$'' equals ``the coset of $L_{Q P}$ with respect to $R_1(S)$.''
Thus, for each $P \in \{I,X,Y,Z\}$, we store the calculated energy in $E^{(i)}_{L_{Q P}}$ in \lineref{line:SA_for_different_reference}.

Note that, in the \algorithmicfor{} loops in Lines \ref{line:for_SA_different_reference} and \ref{line:for_SA_logical_op_different} in \algoref{algo:SA_decoder_different_error_configurations}, each iteration is an independent task, and hence these tasks can trivially be parallelized.
In addition, \lineref{line:determine_logical_operator_class} can be parallelized with Lines \ref{line:C_P} and \ref{line:SA_for_different_reference}.

\begin{publishedalgorithm}[tb]
\caption{The greedy-matching decoder that may produce a different reference configuration}
\label{algo:greedy_matching_different}
\begin{algorithmic}[1]
\Require $S_\bullet$ : Measured syndrome with $\bullet \in \{\mathrm{white}, \mathrm{gray}\}$
\Ensure $C_\bullet$ : Recovery chain, i.e., $\partial{C_\bullet} = S_\bullet$
\Function{greedy\_matching\_diff\_chain}{$S_\bullet$}
\If{$|S_\bullet|$ is odd}
  \State Append the virtual boundary vertex $b$ to $S_\bullet$
\EndIf
\State $K \coloneqq \{ \{u,v\} \,|\,  u,v \in S_\bullet,\, u \neq v\}$  \Comment{\Eqref{eq:list_K}}
\State $C_\bullet \gets \emptyset$

\Statex 

\While{$K$ is not empty}
  \State $w_{\min} \coloneqq  \min\{w(u,v)  \,|\, \{u,v\} \in K \}$
  \State $K_{\min} \coloneqq \{ \{u,v\} \in K \,|\, w(u,v) = w_{\min} \}$  \Comment{\Eqref{eq:list_K_min}}
  \While{$K_{\min}$ is not empty}
    \State Pick a pair $\{u,v\} \in K_{\min}$ uniformly at random.
    \State Add to $C_\bullet$ the edges along a path connecting $u$ and $v$ with the weight $w(u,v)$.
    \State Remove from $K$ and $K_{\min}$ all pairs $\{x,y\}$ such that $x\in\{u,v\}$ or $y\in\{u,v\}$.
  \EndWhile
\EndWhile
\Statex 
\State \Return $C_\bullet$
\EndFunction
\end{algorithmic}

\vspace{\baselineskip}

\begin{algorithmic}[1]
\Require $S \coloneqq (S_\mathrm{white},\, S_\mathrm{gray})$ : Measured syndrome
\Ensure $C$ : Recovery configuration, i.e., $\partial{C} = S$
\Function{greedy\_matching\_different}{$S$}
\State $C_\mathrm{white} \gets$ \Call{greedy\_matching\_diff\_chain}{$S_\mathrm{white}$}
\State $C_\mathrm{gray} \gets$ \Call{greedy\_matching\_diff\_chain}{$S_\mathrm{gray}$}
\State $C \coloneqq (C_\mathrm{white}, C_\mathrm{gray})$
\Statex 
\State \Return $C$
\EndFunction
\end{algorithmic}
\end{publishedalgorithm}

\begin{publishedalgorithm}[tb]
\caption{Decoder using simulated annealing that may start with different reference configurations}
\label{algo:SA_decoder_different_error_configurations}
\begin{algorithmic}[1]
\Require $S$ : Measured syndrome
\Ensure $C$ : Recovery configuration, i.e., $\partial{C} = S$
\Function{SA\_decoder\_different\_reference}{$S$}
\State $R_1 \gets$ \Call{greedy\_matching\_different}{$S$} \Comment{standard reference configuration}  \label{line:standard_error_configuration}
\ForAll {$P \in \{I,X,Y,Z\}$}  \label{line:SA_logical_op_different_init}
   \State $C^{(1)}_{L_P} \coloneqq R_1 \, L_P$  \Comment{standard reference}  \label{line:standard_reference}
   \State $E^{(1)}_{L_P} \gets$ \Call{simulated\_annealing}{\Call{copy}{$C^{(1)}_{L_P}$}}
\EndFor

\Statex 
\For {$i = 2$ to $N_\mathrm{SA}$}  \label{line:for_SA_different_reference}
   \State $R_i \gets$ \Call{greedy\_matching\_different}{$S$}
   \State Determine $Q \in \{I,X,Y,Z\}$ such that $[L_Q] = [R_i \, R_1]$.\\\Comment{e.g., via commutation with $L_X$ and $L_Z$} \label{line:determine_logical_operator_class}
   \ForAll {$P \in \{I,X,Y,Z\}$}  \label{line:for_SA_logical_op_different}
      \State $C^{(i)}_{L_P} \coloneqq R_i \, L_P$  \Comment{Initial configuration for SA}  \label{line:C_P}
      \State $E^{(i)}_{L_{QP}} \gets$ \Call{simulated\_annealing}{$C^{(i)}_{L_P}$}  \label{line:SA_for_different_reference}
   \EndFor
\EndFor
\Statex 
\ForAll {$P \in \{I,X,Y,Z\}$}
   \State $E_{L_P} \coloneqq \min\limits_{i=1,\dots,N_\mathrm{SA}}{E^{(i)}_{L_P}}$ \Comment{estimate of $E^{\min}_{L_P}$}
\EndFor      

\Statex 
\State $P^\ast \coloneqq \argmin\limits_{P \in \{I,X,Y,Z\}}{E_{L_P}}$   \Comment{\Eqref{eq:determine_P_from_E}}
\Statex 
\Statex 
\State \Return $C^{(1)}_{L_{P^\ast}}$  \Comment{defined in \lineref{line:standard_reference}}
\EndFunction
\end{algorithmic}
\end{publishedalgorithm}

\section{Numerical results}
\label{sec:result}

We perform Monte Carlo simulations of our SA decoder under the setting described in \secref{sec:setting}.
In \secref{sec:result_logical_error_rate}, we evaluate the logical error rate of our SA decoder.
In \secref{sec:effect_of_error_configuration}, we examine the validity of our greedy-matching decoder used to construct initial configurations.
We compare elapsed times of our SA decoder with other decoding algorithms in \secref{sec:comparison_time}.

\subsection{Simulation setting}
\label{sec:setting}

In Monte Carlo simulations, we generate samples of true error configurations $\bar{C}$ by means of pseudorandom numbers, drawn from the physical error rate $p$ and the error bias $p_x : p_y : p_z$ we specify.
Performing the stabilizer measurement on the true error configurations $\bar{C}$ yields the syndrome $S$.
Then, we execute each decoder to obtain the recovery configuration $C$.
If the product $\bar{C} \, C$ belongs to a nontrivial logical operator class, i.e., if $[\bar{C} \, C]\in\{[L_X],[L_Y],[L_Z]\}$, then the logical error is said to occur.
By repeating the aforementioned procedure $N_\mathrm{sample}$ times, we estimate the logical error rate $P_\mathrm{L}$.
Except in \secref{sec:comparison_time}, we employ $N_\mathrm{sample}=10^6$ to ensure sufficient statistical precision.
Except for the convergence study in \secref{sec:effect_of_error_configuration}, we refer to \algoref{algo:SA_decoder_different_error_configurations} as \emph{our SA decoder} and employ the number of inverse temperatures $N_\beta = 100$.
The parameters $N_\beta$ and $N_\mathrm{SA}$ are chosen empirically in this work and fixed in advance for each set of simulations, without using an adaptive stopping criterion.

Our SA decoder is applicable to arbitrary biased Pauli noise provided that the decoder is supplied with the error probabilities $p_x,p_y,p_z$ used to define $\mathcal{H}(C)$.
We focus on $Y$-biased noise as a stress test for the MWPM decoder, which neglects $Y$-induced correlations.
For simplicity, we consider the symmetric case $p_x = p_z$.
Under the condition $p_x = p_z$, we can write the bias as $p_x : p_y : p_z = 1 : \eta : 1$ where $\eta$ denotes the $Y$-biased factor.
Since the depolarizing noise is the case where $p_x = p_y = p_z = p/3$, it corresponds to $p_x : p_y : p_z = 1 : 1 : 1$, i.e., $\eta=1$.

\subsection{Logical error rates}
\label{sec:result_logical_error_rate}

Let us compare the decoding accuracy of our SA decoder described in \algoref{algo:SA_decoder_different_error_configurations} with that of a decoder based on IBM ILOG CPLEX Optimizer (CPLEX) described in the Appendix.
Under the dominant-term approximation, the CPLEX decoder directly finds the minimum-energy error configuration consistent with the measured syndrome, whereas the SA decoder minimizes the energy within each logical coset and then selects the coset with the lowest minimum energy.
Thus, for both decoders, the decoding criterion is based on minimum-energy configurations, whereas full maximum-likelihood decoding compares the total weights of the logical cosets.

First, we examine how the logical error rate varies with the code distance $d$.

\begin{figure}[tb]
\includegraphics[width=0.72\linewidth]{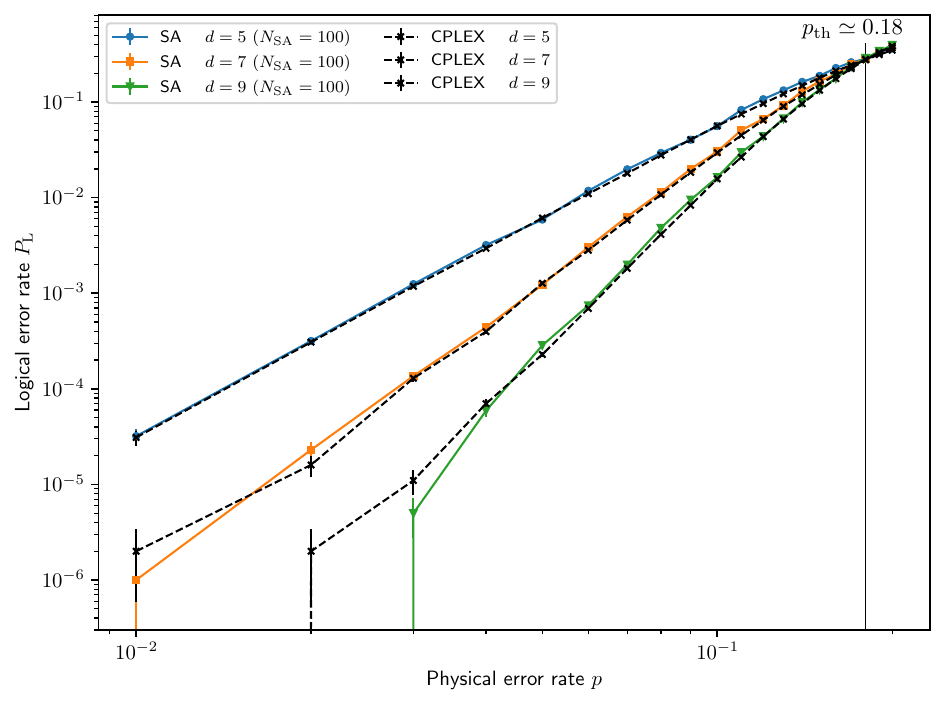}
\caption{Logical error rates under the depolarizing noise for the code distances $d=5,7,9$.
Our SA decoder achieves logical error rates comparable to those of the CPLEX decoder for all $d$.
We employ $N_\mathrm{SA} = 100$, which suffices to reach the logical error rates obtained by the CPLEX decoder.
The error bars indicate one standard deviation.\label{fig:ds_depo}}
\end{figure}

Figure~\ref{fig:ds_depo} depicts the logical error rates $P_\mathrm{L}$ under the depolarizing noise for the code distances $d=5,7,9$.
In \figref{fig:ds_depo}, our SA decoder achieves logical error rates comparable to those of the CPLEX decoder within statistical uncertainty for all $d$ shown.

For both the SA and CPLEX decoders, from the crossing of the finite-distance curves, we estimate a finite-size threshold $p_\mathrm{th}\simeq0.18$.
This estimate is close to the previously reported value $p_\mathrm{th}=0.187$ obtained using an MPS decoder~\cite{xzzx_nature}.
Below the threshold $p_\mathrm{th}$, the logical error rate decreases with increasing $d$, as expected.

\begin{figure}[tb]
\includegraphics[width=0.72\linewidth]{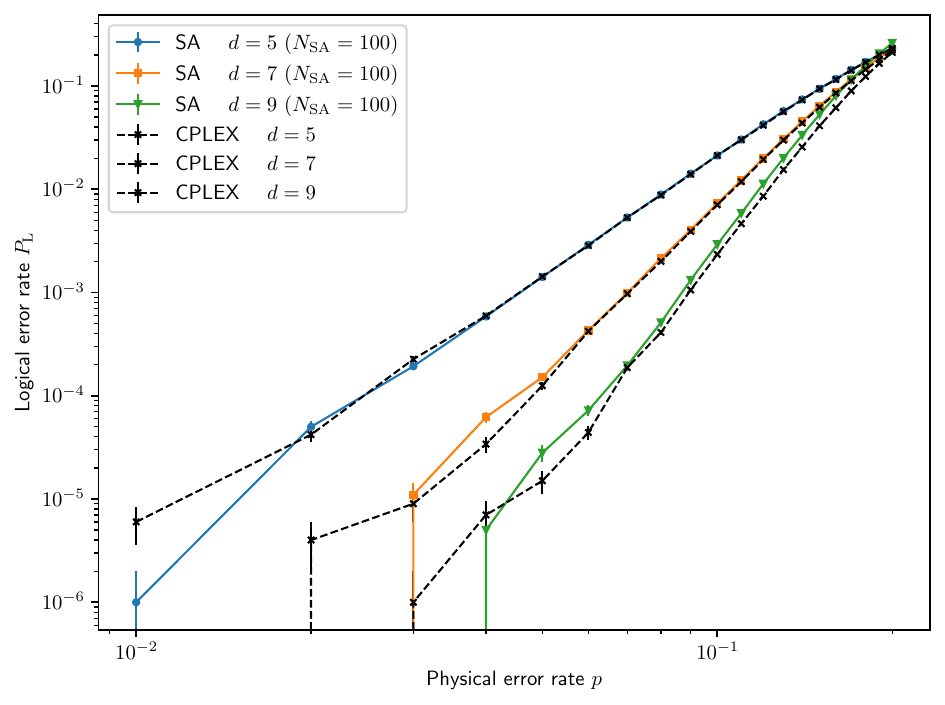}
\caption{Logical error rates under the $Y$-biased noise where $p_x : p_y : p_z = 1 : 5 : 1$ for the code distances $d=5,7,9$.
Our SA decoder achieves logical error rates comparable to those of the CPLEX decoder.
We employ $N_\mathrm{SA} = 100$, which suffices to reach the logical error rates obtained by the CPLEX decoder.
The error bars indicate one standard deviation.\label{fig:ds_py5}}
\end{figure}

Figure~\ref{fig:ds_py5} shows the logical error rates under the $Y$-biased noise where $p_x : p_y : p_z = 1 : 5 : 1$.
In \figref{fig:ds_py5}, for every $d$, our SA decoder achieves logical error rates comparable to those of the CPLEX decoder.
For both SA and CPLEX decoders, the intersection point of these curves lies around $p\simeq0.2$, which we consider to be a finite-size threshold estimate.
Below this threshold, the logical error rate decreases as $d$ increases.

\begin{figure}[tb]
\includegraphics[width=0.72\linewidth]{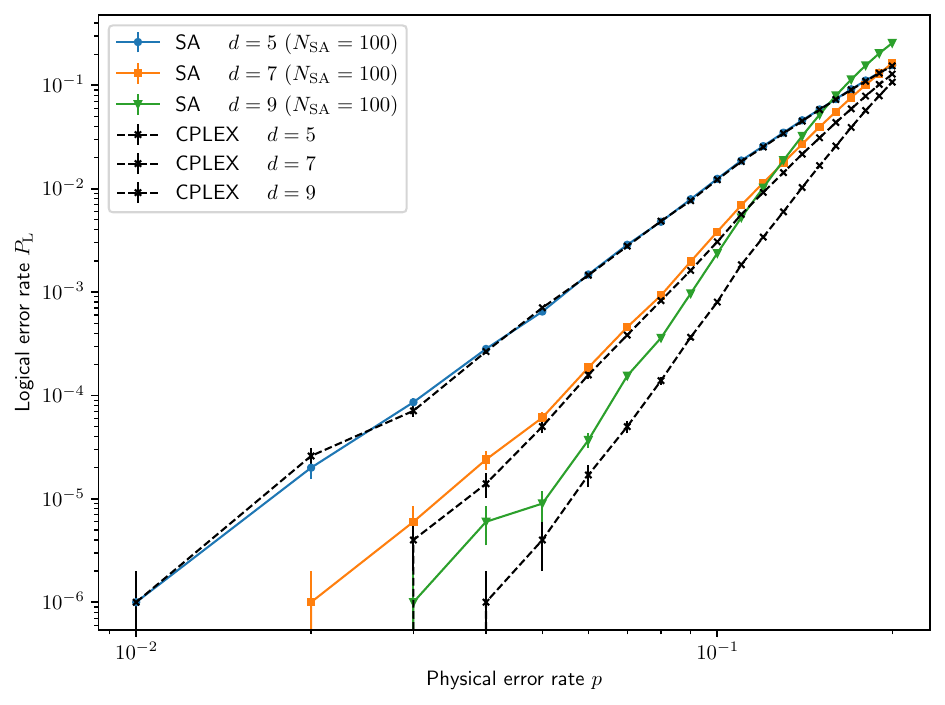}
\caption{Logical error rates under the $Y$-biased noise where $p_x : p_y : p_z = 1 : 10 : 1$ for the code distances $d=5,7,9$.
Our SA decoder does not reach the logical error rate obtained by the CPLEX decoder for $d=9$, even with $N_\mathrm{SA} = 100$.
The error bars indicate one standard deviation.\label{fig:ds_py10}}
\end{figure}

Figure~\ref{fig:ds_py10} shows the logical error rates under the $Y$-biased noise where $p_x : p_y : p_z = 1 : 10 : 1$.
In \figref{fig:ds_py10}, no intersection point for the CPLEX decoder is observed for $d=5,7,9$; hence, we expect that the threshold for the CPLEX decoder lies beyond the plotted range, i.e., $p_\mathrm{th} > 0.2$.
On the other hand, for $d=9$ in \figref{fig:ds_py10}, our SA decoder does not reach the logical error rate obtained by the CPLEX decoder.
The lack of a clear crossing for $d=5,7,9$ may be related to slow mixing of the MCMC method.
For our SA decoder, the curves for $d=7,9$ intersect around $p = 0.11$, which we take as an estimated finite-size threshold.
When $p < 0.11$, the logical error rate decreases as $d$ increases.

Next, we investigate the dependence of the logical error rates $P_\mathrm{L}$ on the number of SA runs, $N_\mathrm{SA}$ for fixed code distance $d=9$.
For comparison, we also calculate the rates by our greedy-matching decoder, i.e., the \texttt{greedy\_matching} function described in \algoref{algo:greedy_matching} and the MWPM decoder, i.e., the \texttt{MWPM} function described in \algoref{algo:MWPM_decoder}.

\begin{figure}[tb]
\includegraphics[width=0.72\linewidth]{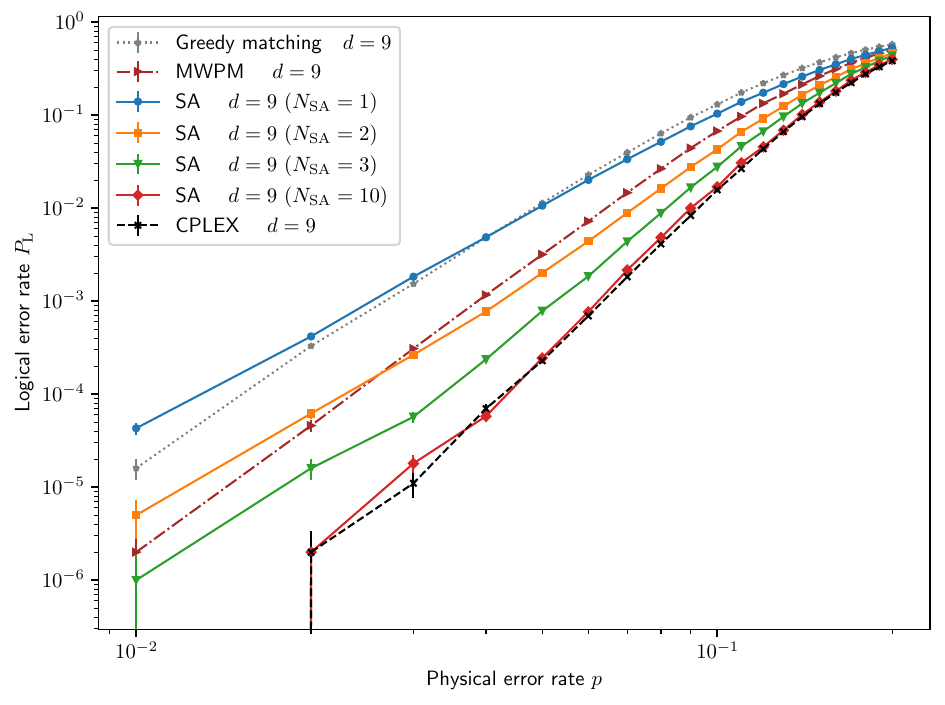}
\caption{Logical error rates under the depolarizing noise and $N_\mathrm{SA}=1,\dots,10$ for the code distance $d=9$.
The error bars indicate one standard deviation.\label{fig:rate_depo}}
\end{figure}

\begin{figure}[tb]
\includegraphics[width=0.72\linewidth]{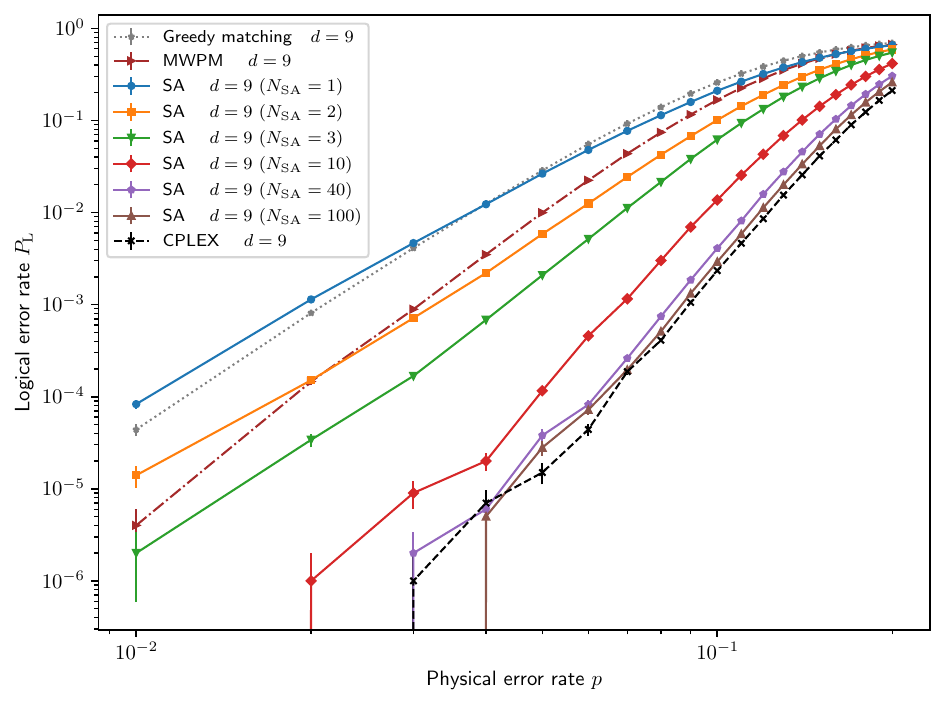}
\caption{Logical error rates under $Y$-biased noise where $p_x : p_y : p_z = 1 : 5 : 1$ and $N_\mathrm{SA}=1,\dots,100$ for the code distance $d=9$.
The error bars indicate one standard deviation.\label{fig:rate_py5}}
\end{figure}

\begin{figure}[tb]
\includegraphics[width=0.72\linewidth]{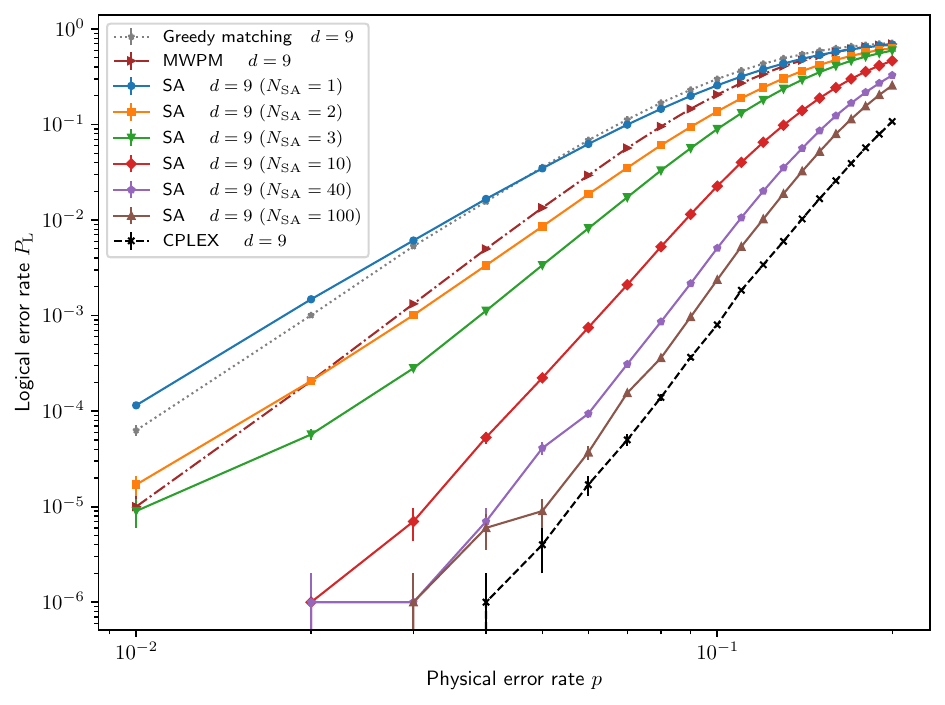}
\caption{Logical error rates under $Y$-biased noise where $p_x : p_y : p_z = 1 : 10 : 1$ and $N_\mathrm{SA}=1,\dots,100$ for the code distance $d=9$.
The error bars indicate one standard deviation.\label{fig:rate_py10}}
\end{figure}

Figure~\ref{fig:rate_depo} shows the logical error rates under the depolarizing noise.
Figures~\ref{fig:rate_py5} and \ref{fig:rate_py10} show the logical error rates under $Y$-biased noises where $p_x : p_y : p_z = 1 : 5 : 1$ and $1 : 10 : 1$ for several values of $N_\mathrm{SA}$, respectively.%

In Figs.\ \ref{fig:rate_depo}--\ref{fig:rate_py10}, we observe that the logical error rate $P_\mathrm{L}$ of our SA decoder decreases monotonically as $N_\mathrm{SA}$ increases.
For $N_\mathrm{SA}\ge 2$, our SA decoder achieves logical error rates lower than our greedy-matching decoder and is therefore an improvement over our greedy-matching.
For $N_\mathrm{SA}\ge 3$, our SA decoder achieves a lower logical error rate than the MWPM decoder.

To reach the accuracy of the CPLEX decoder under the depolarizing noise in \figref{fig:rate_depo}, $N_\mathrm{SA}=10$ is sufficient.
For the noise $p_x : p_y : p_z = 1 : 5 : 1$ in \figref{fig:rate_py5}, $N_\mathrm{SA}=100$ is almost sufficient.
Under the noise $p_x : p_y : p_z = 1 : 10 : 1$ in \figref{fig:rate_py10}, $N_\mathrm{SA}=100$ is not sufficient.
Hence, we observe that the more strongly $Y$-biased the noise is, the larger $N_\mathrm{SA}$ our SA decoder needs to approach the logical error rate obtained by the CPLEX decoder.

\subsection{Effect of initial error configuration on SA performance}
\label{sec:effect_of_error_configuration}

To construct a reference configuration $R(S)$, we consider the following three methods:
\begin{enumerate}
\renewcommand{\labelenumi}{(\arabic{enumi})}
\item \emph{Boundary:} Each syndrome defect is connected to the nearer of the two boundaries (top or bottom for white faces; left or right for gray faces). We call this reference configuration construction routine, instead of the \texttt{greedy\_matching} function in \algoref{algo:SA_decoder}.
\item \emph{Greedy matching (same):} We use \algoref{algo:greedy_matching} as the initializer in \algoref{algo:SA_decoder}; it returns the same reference configuration on every call.
\item \emph{Greedy matching (different):} It is \algoref{algo:greedy_matching_different} invoked in the SA decoder of \algoref{algo:SA_decoder_different_error_configurations}, described in \secref{sec:SA_decoder_greedy_matching_with_different_distance}. For every invocation, it may produce a different reference configuration. We use this method in this paper, unless otherwise stated.
\end{enumerate}

To compare the effect of these methods on the convergence of our SA decoder, we evaluate the ratio of logical error rates
\begin{equation}
R_\mathrm{L} \coloneqq \left( \dfrac{P^{\mathrm{(SA)}}_\mathrm{L}}{P^{\mathrm{(CPLEX)}}_\mathrm{L}} \right)^{\frac{2}{d+1}}
,
\label{eq:ratio_of_logical_error_rate_to_CPLEX}
\end{equation}
where $P^\mathrm{(SA)}_\mathrm{L}$ and $P^\mathrm{(CPLEX)}_\mathrm{L}$ are the logical error rates obtained by the SA and CPLEX decoders, respectively.
Here, the exponent $2/(d+1)$ is introduced to cancel out $d$ dependence.
A value $R_\mathrm{L} = 1$ indicates that the SA decoder matches the CPLEX benchmark.
Values above $1$ indicate worse performance than CPLEX.

\begin{figure}[tb]
\centering
\sidecaption{subfig_depo:a}\raisebox{-\height}{\includegraphics[width=0.38\linewidth]{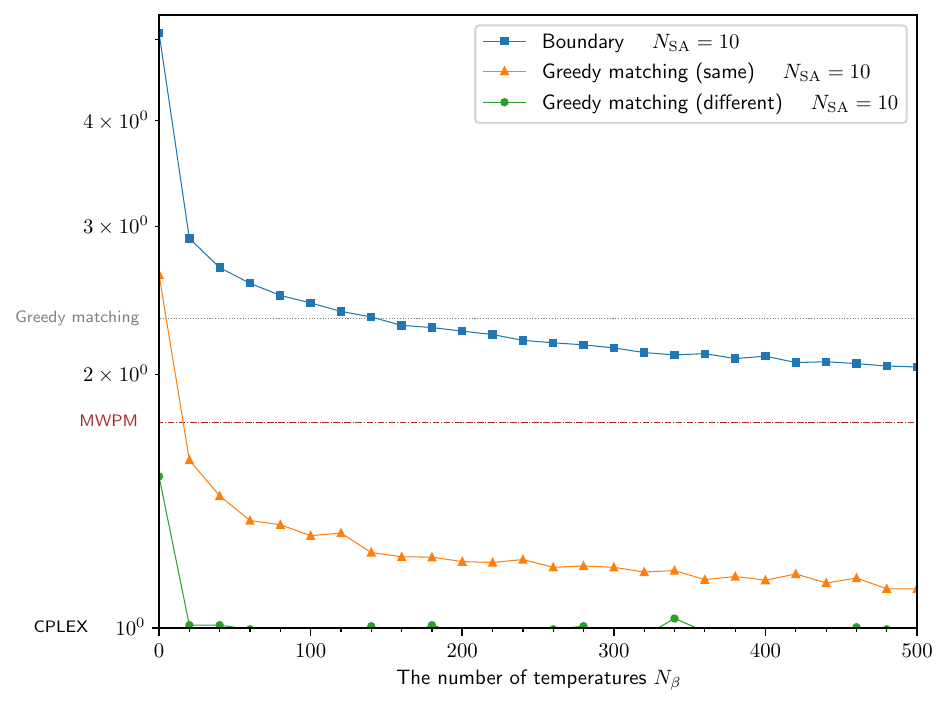}}%
\sidecaption{subfig_depo:b}\raisebox{-\height}{\includegraphics[width=0.38\linewidth]{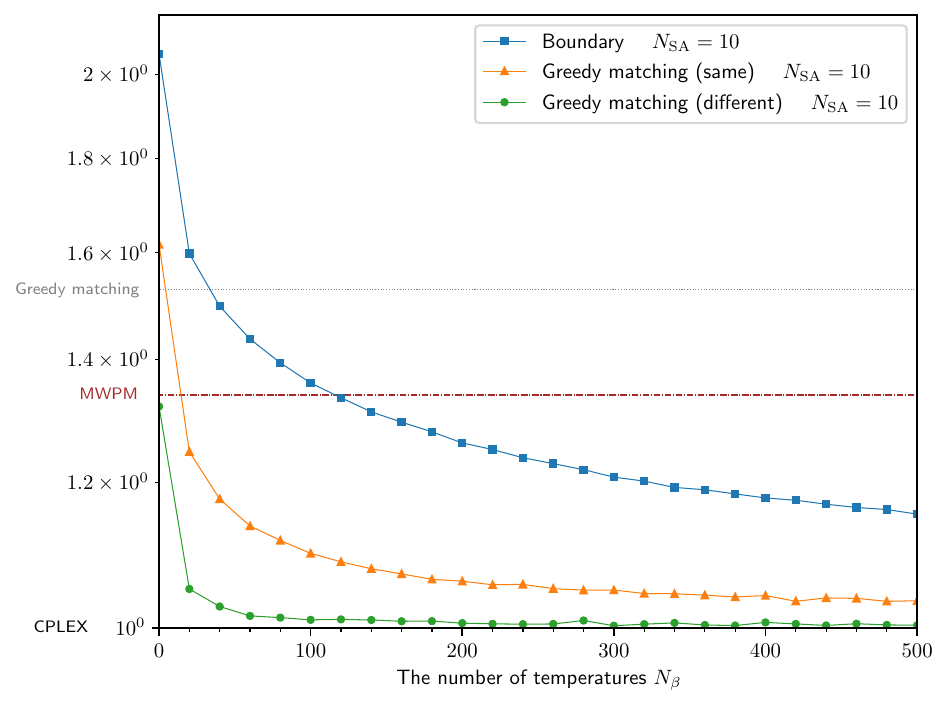}}%
\hfill
\sidecaption{subfig_depo:c}\raisebox{-\height}{\includegraphics[width=0.38\linewidth]{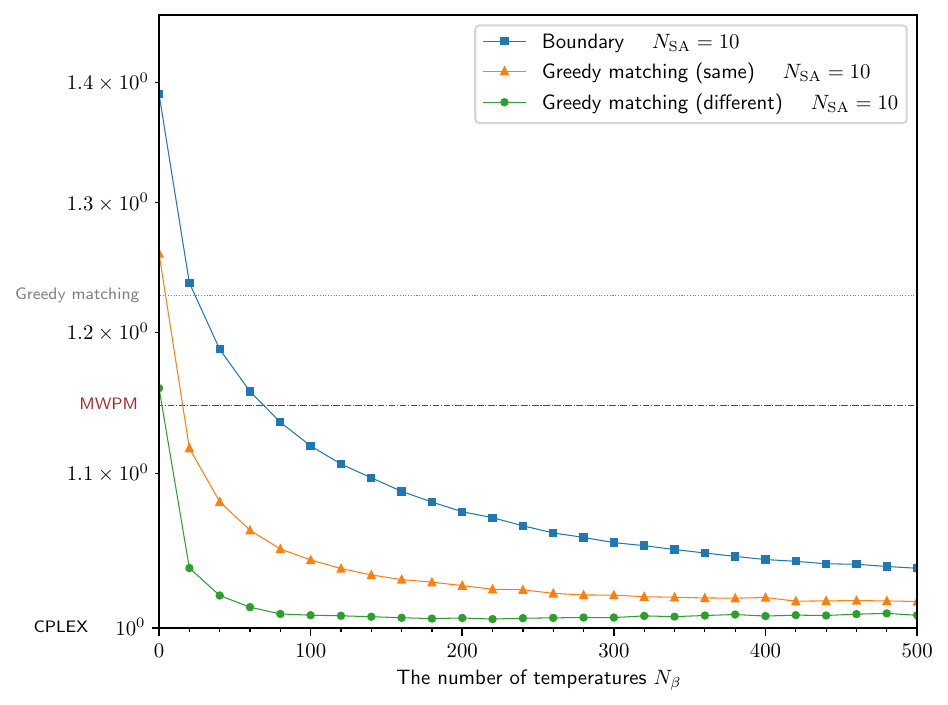}}%
\caption{Convergence of our SA decoder for different construction methods of reference configurations under the depolarizing noise.
The horizontal axis is the number of inverse temperatures $N_\beta$, and the vertical axis is the ratio $R_\mathrm{L}$.
We employ $d=9$ and $N_\mathrm{SA}=10$.
(a) $p=0.04$, (b) $p=0.1$, and (c) $p=0.15$.
Dotted gray and dash-dotted brown horizontal lines show the ratios obtained by the greedy-matching and MWPM decoders, respectively.
\label{fig:conv_error_configuration_depo}}
\end{figure}

\begin{figure}[tb]
\centering
\sidecaption{subfig_biased5:a}\raisebox{-\height}{\includegraphics[width=0.38\linewidth]{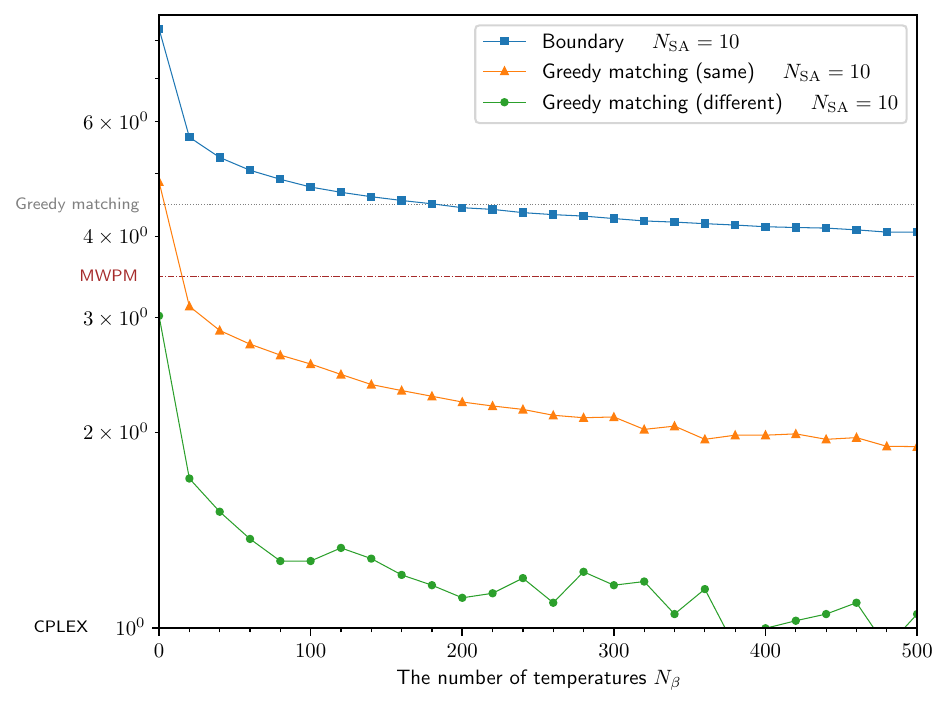}}%
\sidecaption{subfig_biased5:b}\raisebox{-\height}{\includegraphics[width=0.38\linewidth]{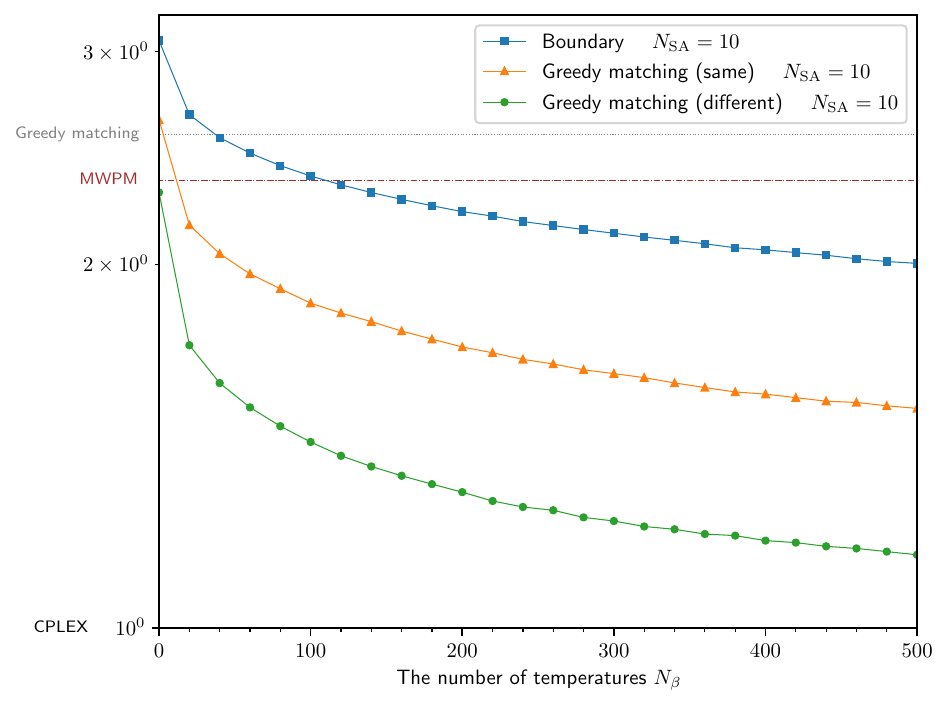}}%
\hfill
\sidecaption{subfig_biased5:c}\raisebox{-\height}{\includegraphics[width=0.38\linewidth]{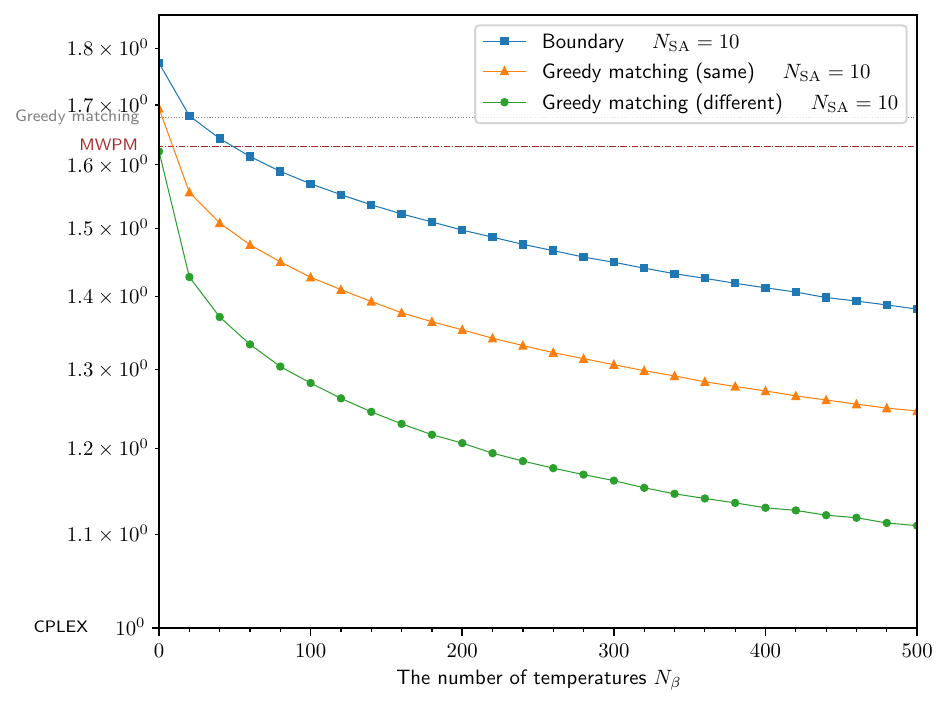}}%
\caption{Convergence of our SA decoder for different construction methods of reference configurations under the $Y$-biased noise such that $p_x : p_y : p_z = 1 : 5 : 1$.
The horizontal axis is the number of inverse temperatures $N_\beta$, and the vertical axis is the ratio $R_{\mathrm{L}}$.
We employ $d=9$ and $N_\mathrm{SA}=10$.
(a) $p=0.04$, (b) $p=0.1$, and (c) $p=0.15$.
Dotted gray and dash-dotted brown horizontal lines show the ratios obtained by the greedy-matching and MWPM decoders, respectively.
\label{fig:conv_error_configuration_biased1_5_1}}
\end{figure}

\begin{figure}[tb]
\centering
\sidecaption{subfig_biased10:a}\raisebox{-\height}{\includegraphics[width=0.38\linewidth]{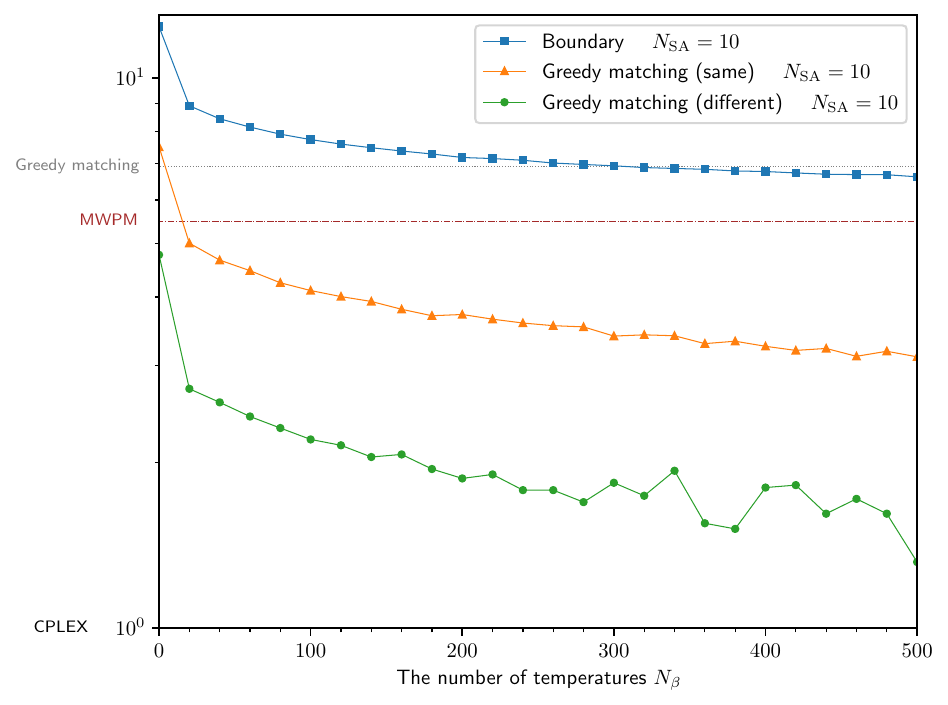}}%
\sidecaption{subfig_biased10:b}\raisebox{-\height}{\includegraphics[width=0.38\linewidth]{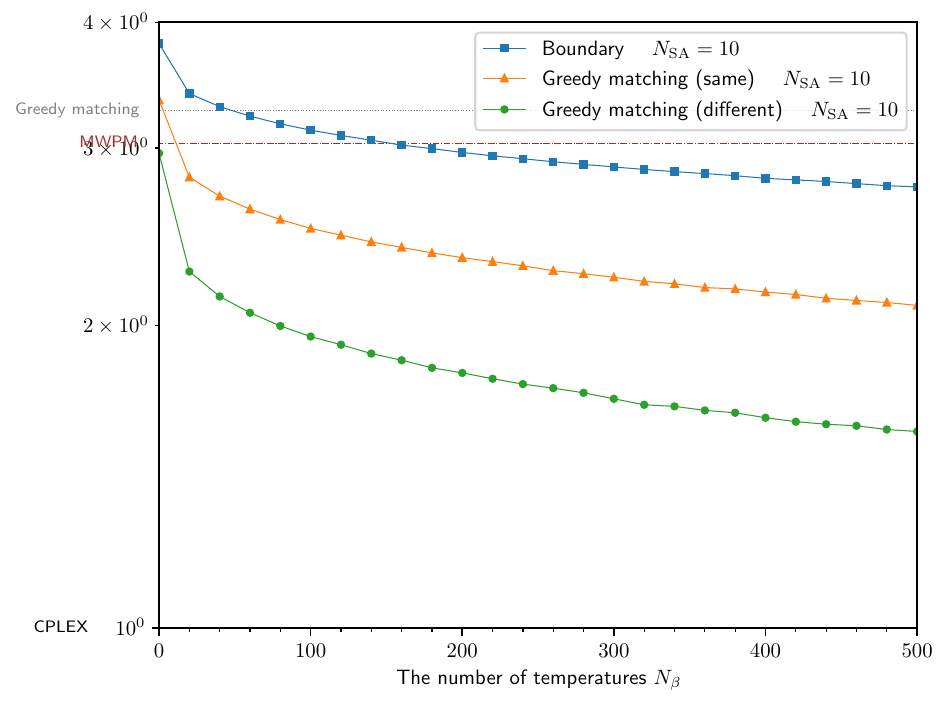}}%
\hfill
\sidecaption{subfig_biased10:c}\raisebox{-\height}{\includegraphics[width=0.38\linewidth]{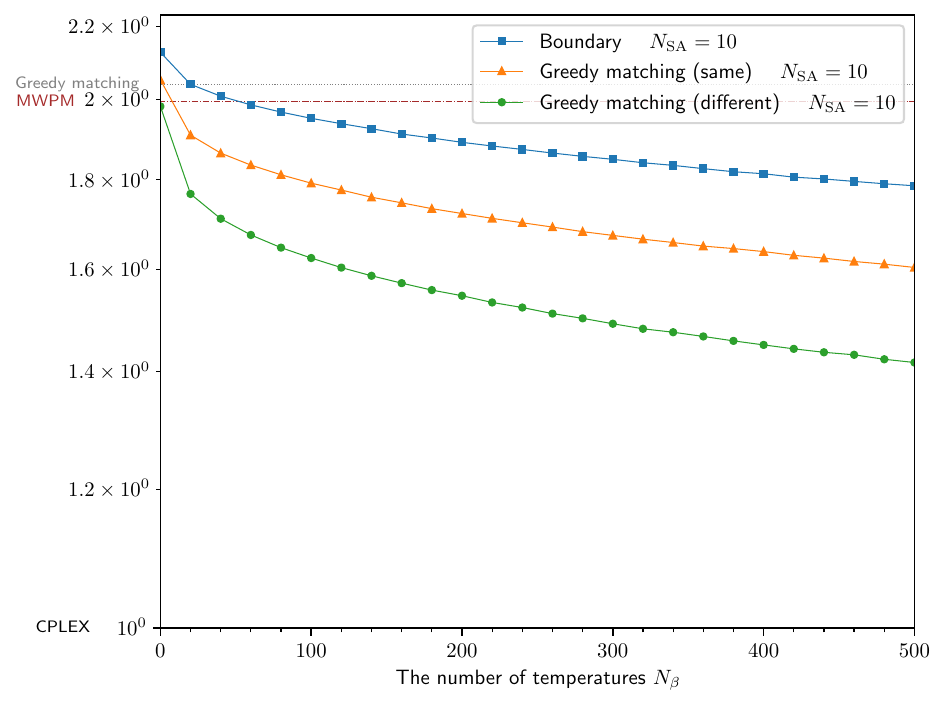}}%
\caption{Convergence of our SA decoder for different construction methods of reference configurations under the $Y$-biased noise such that $p_x : p_y : p_z = 1 : 10 : 1$.
The horizontal axis is the number of inverse temperatures $N_\beta$, and the vertical axis is the ratio $R_\mathrm{L}$.
We employ $d=9$ and $N_\mathrm{SA}=10$.
(a) $p=0.04$, (b) $p=0.1$, and (c) $p=0.15$.
Dotted gray and dash-dotted brown horizontal lines show the ratios obtained by the greedy-matching and MWPM decoders, respectively.
\label{fig:conv_error_configuration_biased1_10_1}}
\end{figure}

Figure~\ref{fig:conv_error_configuration_depo} shows the ratio $R_{\mathrm{L}}$ for varying the number of inverse temperatures $N_\beta$ under the depolarizing noise.
Similarly, Figs.~\ref{fig:conv_error_configuration_biased1_5_1} and \ref{fig:conv_error_configuration_biased1_10_1} show the ratio $R_{\mathrm{L}}$ under the $Y$-biased noises where $p_x : p_y : p_z = 1 : 5 : 1$ and $1 : 10 : 1$, respectively.
In Figs.\ \ref{fig:conv_error_configuration_depo}--\ref{fig:conv_error_configuration_biased1_10_1}, the ratio values $R_\mathrm{L}$ obtained by the greedy-matching and MWPM decoders are drawn by dotted gray and dash-dotted brown horizontal lines, respectively.%

In Figs.\ \ref{fig:conv_error_configuration_depo}--\ref{fig:conv_error_configuration_biased1_10_1}, under all the noises and for all the reference configuration construction methods, we see that the ratio $R_{\mathrm{L}}$ broadly decreases to the value $1$ as $N_\beta$ increases.
For $N_\beta \ge 20$, the ``Greedy matching (same)'' and ``Greedy matching (different)'' methods outperform the MWPM decoder.

For the ``Greedy matching (same)'' and ``Greedy matching (different)'', when $N_\beta=0$, the \texttt{simulated\_annealing} function performs no Metropolis updates, so the decoder compares only the energies of the initial configurations.
We refer to these variants as ``initial-configuration-only'' variants.
In all the cases (a)--(c) of Figs.\ \ref{fig:conv_error_configuration_depo}--\ref{fig:conv_error_configuration_biased1_10_1}, the initial-configuration-only variant of the ``Greedy matching (different)'' method achieves higher accuracy than the MWPM decoder.

In Figs.\ \ref{fig:conv_error_configuration_depo}--\ref{fig:conv_error_configuration_biased1_10_1}, for all plotted values of $N_\beta$ and for all noise biases, ``Greedy matching (different)'' consistently yields the lowest $R_\mathrm{L}$ among the three methods, followed by ``Greedy matching (same)'', while the ``Boundary'' method performs worst.
We attribute these accuracy differences to the following factors:
\begin{enumerate}
\renewcommand{\labelenumi}{(\arabic{enumi})}
\item The ``Boundary'' method ignores the error probabilities $p_x$, $p_y$, and $p_z$, whereas ``Greedy matching (same)'' incorporates $p_x$ and $p_z$ through the weights $\tilde\alpha_x$ and $\tilde\alpha_z$.
As a result, ``Greedy matching (same)'' yields a bias-informed starting configuration.
\item The ``Greedy matching (different)'' method randomizes the tie breaking among equal-distance pairs, generating diverse initial configurations across runs.
This diversity increases the chance of reaching lower-energy configurations and thereby reduces the logical error rate.
\end{enumerate}

\subsection{Comparison of decoding time with other decoding algorithms}
\label{sec:comparison_time}

We now evaluate the decoding time of our SA decoder.

\begin{figure}[tb]
\centering
\sidecaption{time_depo:a}\raisebox{-\height}{\includegraphics[width=0.38\linewidth]{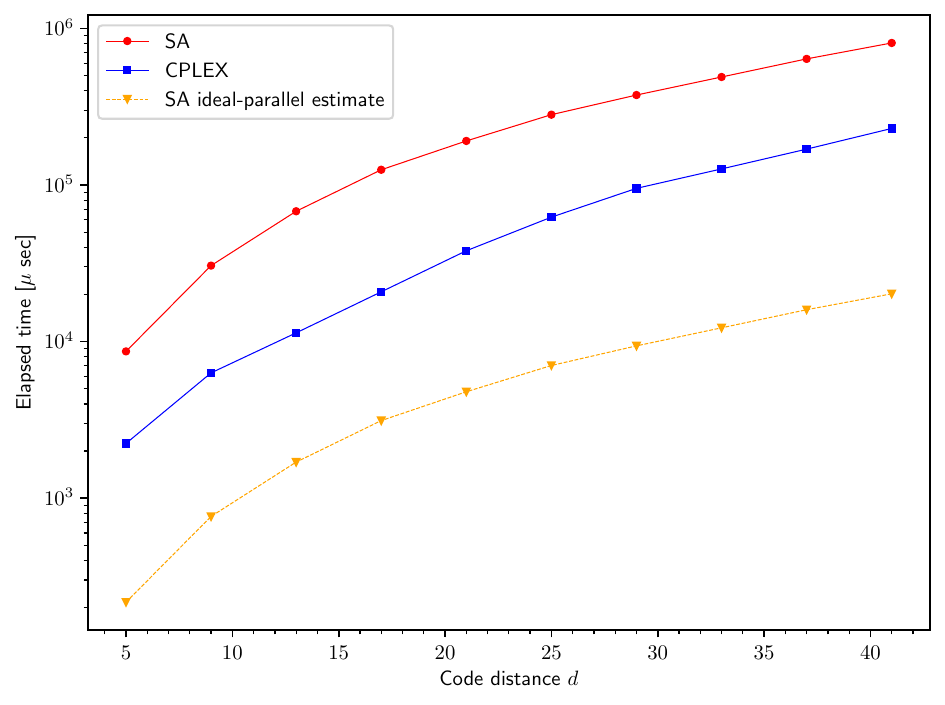}}%
\hfill
\sidecaption{time_depo:b}\raisebox{-\height}{\includegraphics[width=0.38\linewidth]{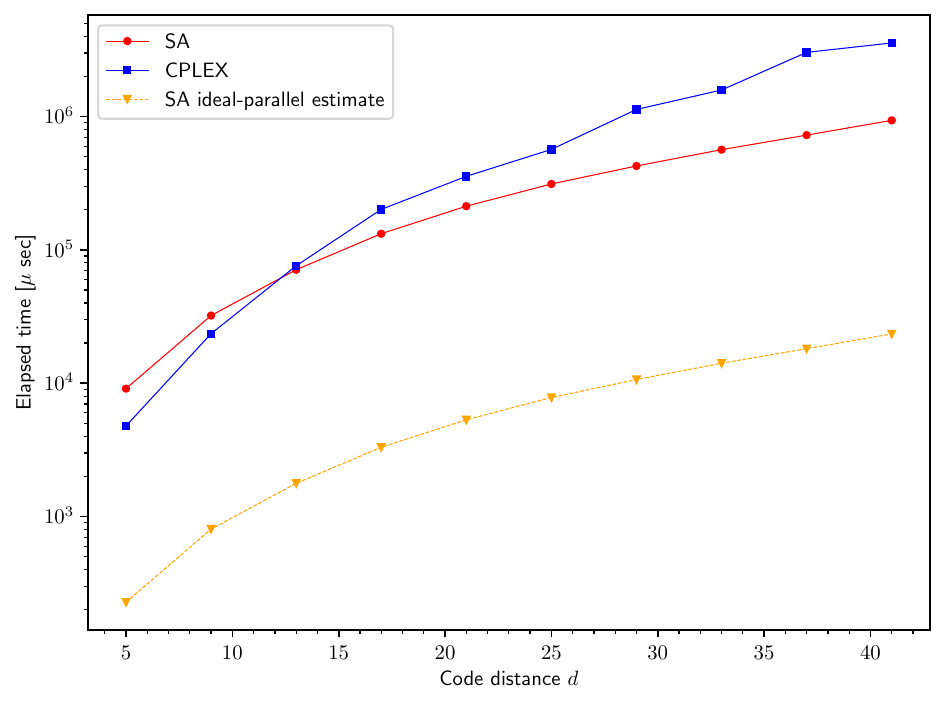}}
\sidecaption{time_depo:c}\raisebox{-\height}{\includegraphics[width=0.38\linewidth]{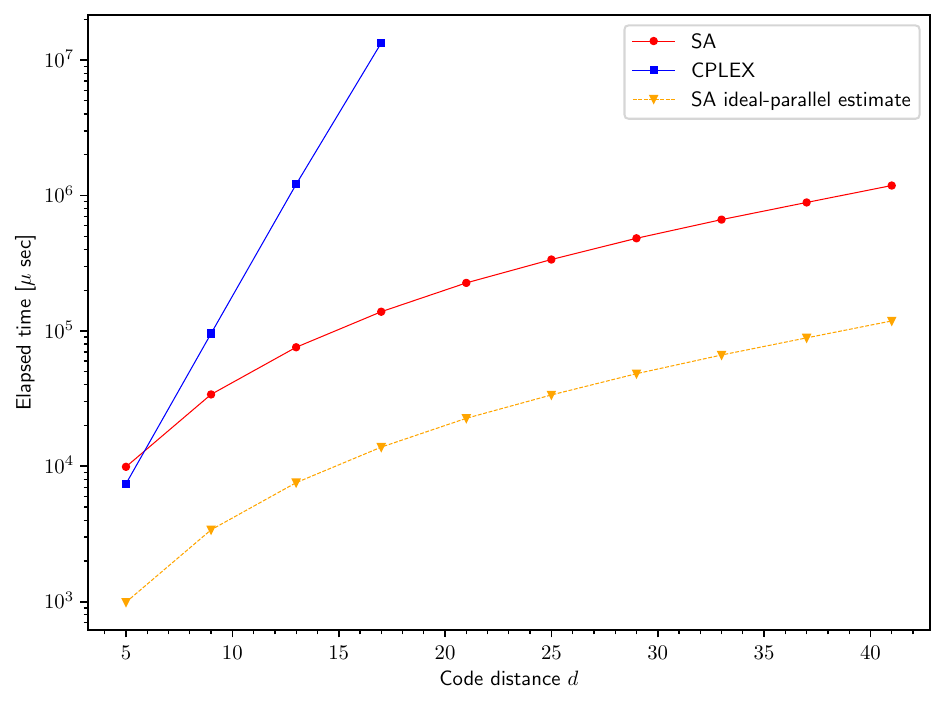}}
\caption{Elapsed times for the SA and CPLEX decoders under the depolarizing noise.
The vertical axis is on a logarithmic scale.
The orange dashed line shows an idealized parallel-run-time estimate obtained by dividing the sequential SA time by $4N_\mathrm{SA}$; it is not a measured parallel run-time.
(a) $p=0.02$, (b) $p=0.1$, and (c) $p=0.15$.
In panel (c), the CPLEX decoder exhibits exponential scaling for the code distance $d$.
\label{fig:time_SA_CPLEX_depo}}
\end{figure}

\begin{figure}[tb]
\begin{center}
\centering
\sidecaption{time_py5:a}\raisebox{-\height}{\includegraphics[width=0.38\linewidth]{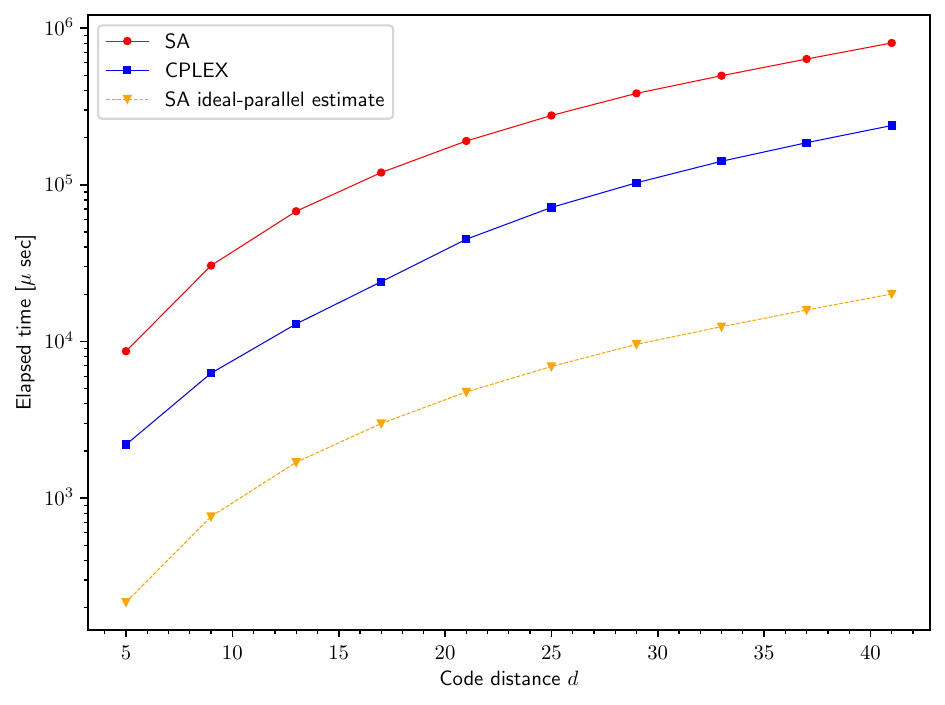}}%
\hfill
\sidecaption{time_py5:b}\raisebox{-\height}{\includegraphics[width=0.38\linewidth]{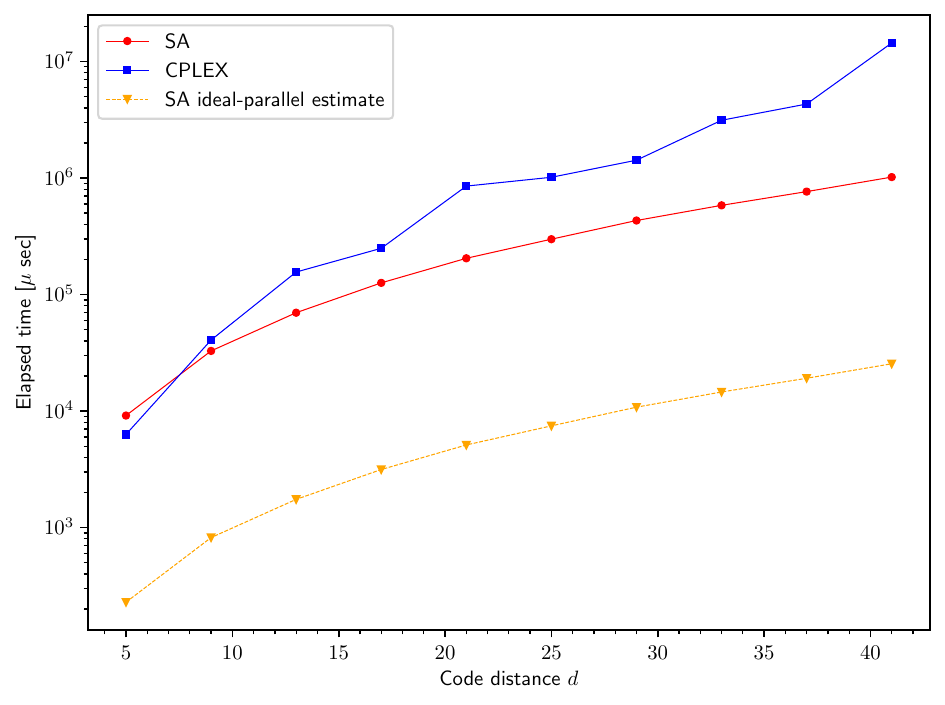}}
\sidecaption{time_py5:c}\raisebox{-\height}{\includegraphics[width=0.38\linewidth]{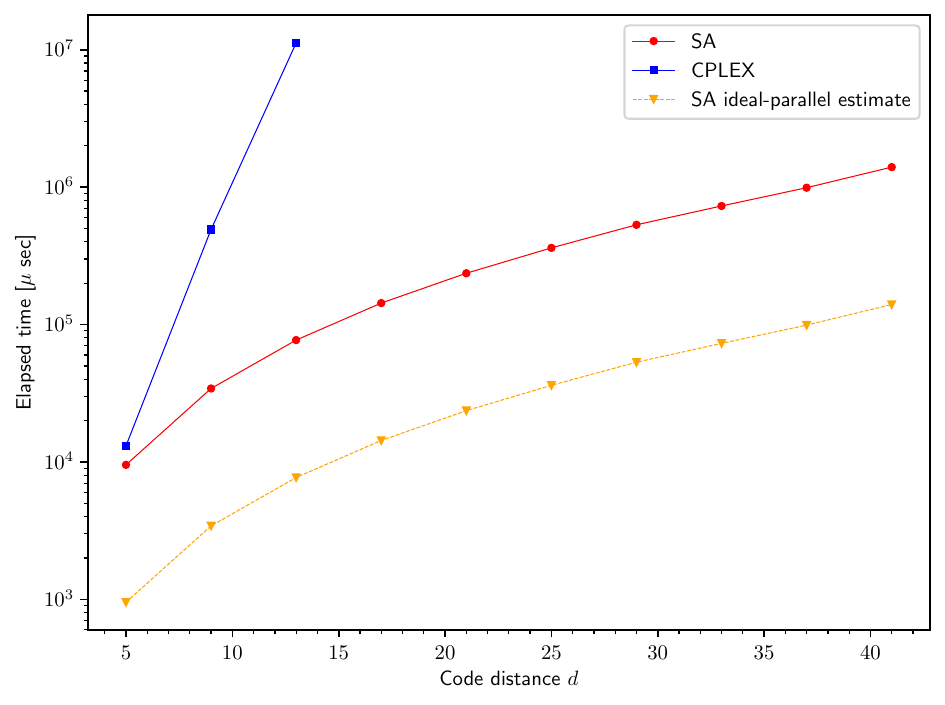}}
\end{center}
\caption{Elapsed times for the SA and CPLEX decoders under the $Y$-biased noise where $p_x : p_y : p_z = 1 : 5 : 1$.
The vertical axis is on a logarithmic scale.
The orange dashed line shows an idealized parallel-run-time estimate obtained by dividing the sequential SA time by $4N_\mathrm{SA}$; it is not a measured parallel run-time.
(a) $p=0.02$, (b) $p=0.1$, and (c) $p=0.15$.
In panel (c), the CPLEX decoder exhibits exponential scaling for the code distance $d$.
\label{fig:time_SA_CPLEX_py5}}
\end{figure}

\begin{figure}[tb]
\centering
\sidecaption{time_py10:a}\raisebox{-\height}{\includegraphics[width=0.38\linewidth]{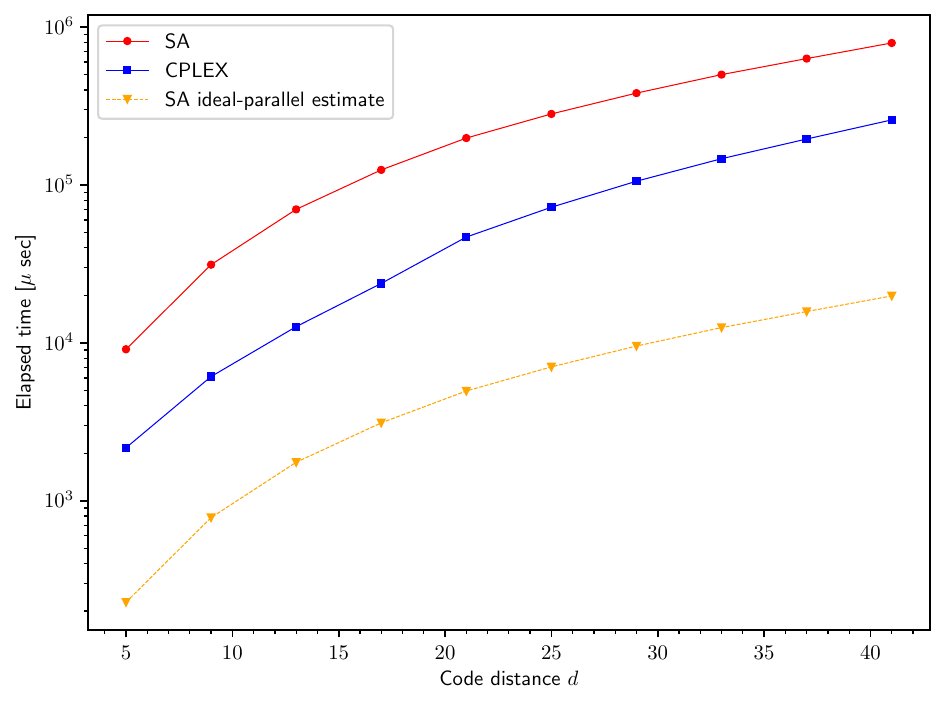}}%
\hfill
\sidecaption{time_py10:b}\raisebox{-\height}{\includegraphics[width=0.38\linewidth]{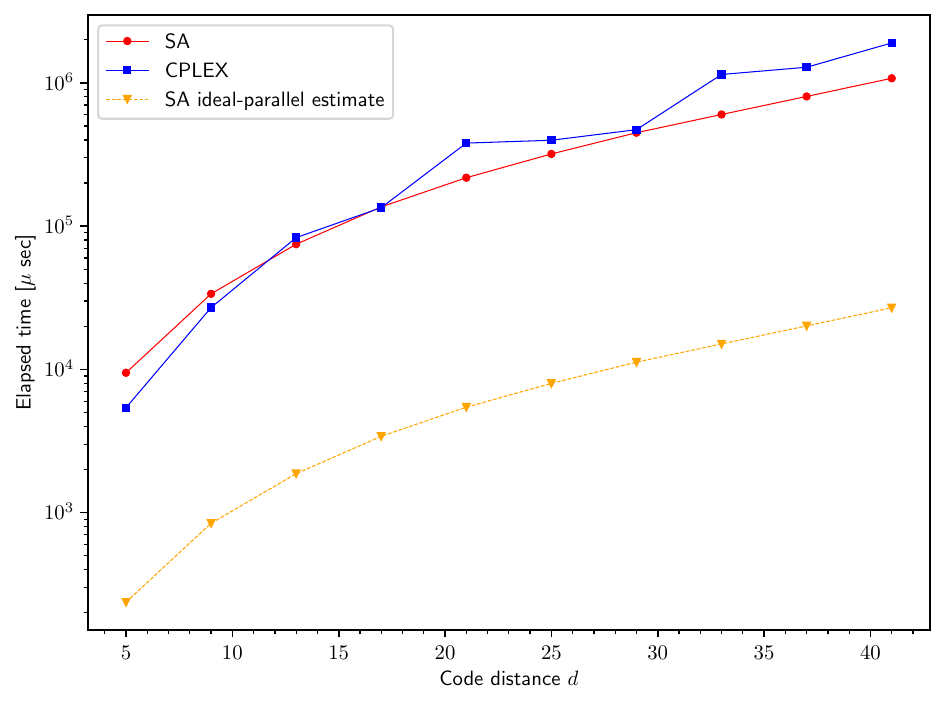}}
\sidecaption{time_py10:c}\raisebox{-\height}{\includegraphics[width=0.38\linewidth]{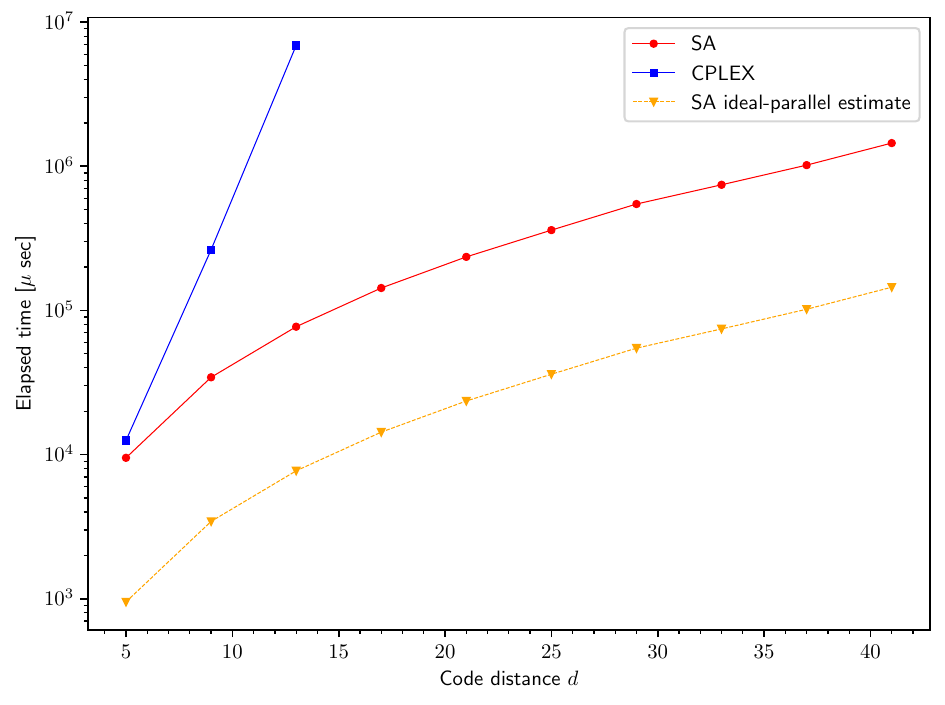}}
\caption{Elapsed times for the SA and CPLEX decoders under the $Y$-biased noise where $p_x : p_y : p_z = 1 : 10 : 1$.
The vertical axis is on a logarithmic scale.
The orange dashed line shows an idealized parallel-run-time estimate obtained by dividing the sequential SA time by $4N_\mathrm{SA}$; it is not a measured parallel run-time.
(a) $p=0.02$, (b) $p=0.1$, and (c) $p=0.15$.
In panel (c), the CPLEX decoder exhibits exponential scaling for the code distance $d$.
\label{fig:time_SA_CPLEX_py10}}
\end{figure}

First, \figref{fig:time_SA_CPLEX_depo} shows the elapsed times for the SA and CPLEX decoders for various code distances $d$ under the depolarizing noise.
Similarly, Figs.\ \ref{fig:time_SA_CPLEX_py5} and \ref{fig:time_SA_CPLEX_py10} show the elapsed times under $Y$-biased noises where $p_x : p_y : p_z = 1 : 5 : 1$ and $1 : 10 : 1$, respectively.
For measuring the elapsed times, we execute the SA and CPLEX decoders without any parallelization.
From Figs.\ \ref{fig:time_SA_CPLEX_depo}--\ref{fig:time_SA_CPLEX_py10}, we see that the larger $d$ is, the longer both the SA and CPLEX decoders take.

For each decoder, we examine the dependence on $d$ and $p$.
For all the physical error rates $p$, our SA decoder takes approximately the same time, because the iteration count is determined solely by $N_\beta$ and $d$ and does not depend on the syndrome weight $|S|$.
Recalling that for each inverse temperature, we perform the Metropolis step $\mathcal{O}(N_\mathrm{stabilizer}) = \mathcal{O}(d^2)$ times, the total elapsed time scales accordingly.
On the other hand, the CPLEX decoder exhibits exponential scaling in computation time with respect to the code distance $d$.
This exponential behavior is clearly visible for case (c) $p=0.15$ in Figs.\ \ref{fig:time_SA_CPLEX_depo}--\ref{fig:time_SA_CPLEX_py10}, where it is prohibitive to calculate for $d > 11$.

Let us now compare the elapsed times of our SA decoder and the CPLEX decoder.
For $p=0.15$, our SA decoder is faster than the CPLEX decoder in the measured sequential implementation.
For $p=0.02$ and $p=0.1$, the measured sequential SA run-time is longer than the CPLEX run-time.
This indicates that improving the sequential run-time in the low-$p$ regime remains an important task.

In each of Figs.\ \ref{fig:time_SA_CPLEX_depo}--\ref{fig:time_SA_CPLEX_py10}, the orange dashed line labeled ``SA ideal-parallel estimate'' is an idealized parallel-run-time estimate obtained by dividing the sequential SA time by $4N_\mathrm{SA}=40$.
This curve is not a measured parallel run-time; it assumes perfect parallel efficiency, with parallel overhead neglected.
In these figures, comparison of this estimate with the CPLEX time suggests possible run-time competitiveness under these idealized assumptions.

\begin{figure}[tb]
\centering
\includegraphics[width=0.72\linewidth]{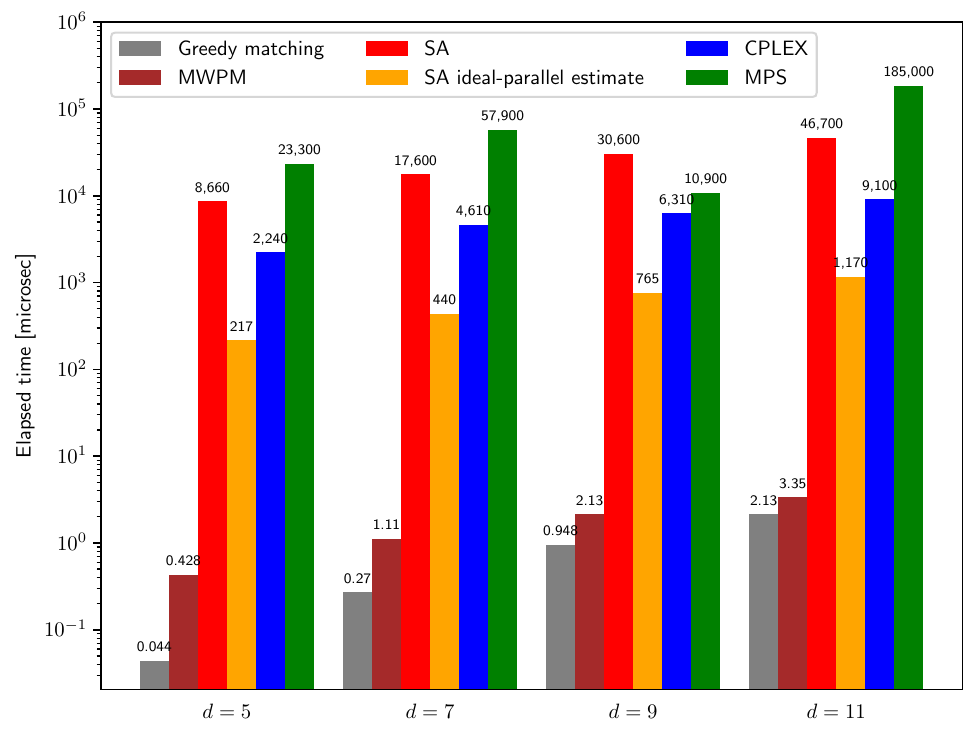}
\caption{Elapsed times (microseconds) of various decoding algorithms.
Each value is an average over $N_\mathrm{sample}$ true error configurations, where $N_\mathrm{sample} = 10^6$ for the greedy-matching and MWPM decoders, and $N_\mathrm{sample} = 100$ for the SA, CPLEX, and MPS decoders due to their higher per-instance cost.
We employ the depolarizing noise with the physical error rate $p = 0.02$.
Here, the label ``SA'' indicates ``Greedy matching (different)'' type with the parameters $N_\beta = 100$ and $N_\mathrm{SA} = 10$.
The label ``SA ideal-parallel estimate'' denotes an idealized parallel-run-time estimate obtained by dividing the sequential SA time by $4 N_\mathrm{SA} = 40$; it is not a measured parallel run-time.
The MPS timing is shown only as an indicative run-time reference, not as a controlled run-time benchmark, because the MPS timing was obtained using a publicly available Python implementation of an MPS decoder for the CSS surface code, rather than using an XZZX-specific implementation.
The ``Greedy matching'' and ``MWPM'' correspond to Algorithms~\ref{algo:greedy_matching} and \ref{algo:MWPM_decoder}, respectively.\label{fig:elapsed_time}}
\end{figure}

Next, we present the elapsed times for various decoding algorithms in \figref{fig:elapsed_time}.
For our SA decoder, we set the number of inverse temperatures to $N_\beta = 100$ and the number of SA runs to $N_\mathrm{SA} = 10$ in \algoref{algo:SA_decoder_different_error_configurations}.
Note that increasing $N_\mathrm{SA}$ increases the total cost roughly linearly, since each SA run has a similar per-run cost.

For the MPS decoder, we use the CSS surface code implementation in Ref.~\cite{qecsim} with truncation dimension $\chi=8$, because the XZZX-specific implementation used in Ref.~\cite{xzzx_nature} is not publicly available.
Accordingly, we do not use this implementation for an accuracy comparison, and its timing in Fig.~\ref{fig:elapsed_time} is included only as an indicative run-time reference rather than as a controlled benchmark against our SA implementation.
In particular, the MPS timing is obtained from a Python implementation for the CSS surface code, whereas our SA decoder is implemented in C++.
For fixed $\chi$, the MPS decoder scales as $\mathcal{O}(N_\mathrm{q})=\mathcal{O}(d^2)$, up to polynomial factors in $\chi$~\cite{efficient_MLD,tuckett_tailoring}.
For fixed $N_\beta$ and $N_\mathrm{SA}$, our SA decoder also scales as $\mathcal{O}(d^2)$ in $d$, although constant factors, implementation details, and parameter dependencies differ.
A controlled comparison with an XZZX-specific MPS implementation is left for future work.

Our SA decoder labeled by ``SA'' is slower than the CPLEX decoder in this sequential implementation.
Nevertheless, the corresponding ``SA ideal-parallel estimate'', obtained by dividing the ``SA'' time by $4 N_\mathrm{SA}$, suggests possible run-time competitiveness with the CPLEX decoder under the assumption of perfect scaling.
This estimate is idealized and should not be regarded as a measured parallel run-time.
Quantifying the achievable parallel efficiency and overhead is left for future work.

From \figref{fig:elapsed_time}, we observe that our greedy-matching decoder is faster than the MWPM decoder.
This result is consistent with the fact that the worst-case computational complexity of our greedy-matching decoder is $\mathcal{O}(|S_\bullet|^2 \log{|S_\bullet|^2}) = \mathcal{O}(d^4 \log{d})$, as discussed in \secref{sec:greedy_matching}, which is lower than the worst-case complexity of the MWPM decoder $\mathcal{O}(|S_\bullet|^3 \log{|S_\bullet|}) = \mathcal{O}(d^6 \log{d})$ in Ref.~\cite{BlossomV}
\footnote{Note that, in circuit-level noise model, the MWPM decoder is used with the Dijkstra algorithm to find a path with a minimum weight between any possible pair of syndrome defects, whereas in the code capacity noise model, the Dijkstra algorithm is not needed, because the weights are explicitly given by Eqs.~\eqref{eq:distance_u_v_internal} and \eqref{eq:distance_u_v_virtual_white}--\eqref{eq:distance_u_boundary}.}.
We emphasize that practical MWPM decoders often show much better average-case scaling [e.g., $\mathcal{O}(d^2)$ reported in Ref.~\cite{MWPM_complexity}], and a similar empirical study for our greedy-matching decoder is left for future work.

\section{Discussion}
\label{sec:discussion}

As in \secref{sec:effect_of_error_configuration}, under much more strongly $Y$-biased noise, our SA decoder requires a large $N_\mathrm{SA}$ in order to approach the logical error rate obtained by the CPLEX decoder.
A possible remedy is to modify the definition of stabilizer generators, analogously to the tailored code for the CSS surface code~\cite{tuckett_tailoring}.
For example, in regimes where $p_y$ dominates (e.g., $p_y \ge p_z \ge p_x$), one could consider an ``XYYX'' variant whose stabilizer generators include $Y$ operators.
This modified code can detect the end points of error chains consisting of $Z$ errors on horizontal edges and $Y$ errors on vertical edges.
Thus, in our greedy-matching algorithm, we can take into account the probabilities $p_z$ and $p_y$.
We therefore expect to obtain a reference configuration, closely resembling the true error configuration.
By employing this better initial configuration, our SA decoder is expected to converge more quickly toward the logical error rate obtained by the CPLEX decoder.
Such a treatment can be applied to accommodate the given noise bias.

A key advantage of our SA decoder is that it provides explicit control over the trade-off between computational cost and accuracy by adjusting $N_\beta$.
That is, increasing the number of inverse temperatures $N_\beta$ typically improves the decoding accuracy, but also increases the computational cost.
In contrast, the cost of deterministic algorithms such as the MWPM, CPLEX, and MPS decoders cannot be easily adjusted, and their parallelization is neither straightforward nor efficient.

\section{Conclusion}
\label{sec:conclusion}

We have formulated and demonstrated our SA decoder for the XZZX code.
As an initial configuration for SA, we have proposed to employ recovery chains obtained by our decoder that utilizes a greedy-matching graph algorithm.

We assumed the code capacity noise model.
Under this model, we performed numerical simulations under depolarizing noise and $Y$-biased Pauli noise.

We observed that our SA decoder is more accurate than the MWPM decoder, which cannot take into account the probability of $Y$ errors.
Moreover, under depolarizing noise and $Y$-biased noise with $p_x:p_y:p_z=1:5:1$, our SA decoder achieves logical error rates $P_\mathrm{L}$ comparable to those of the CPLEX decoder for all code distances $d=5,7,9$.
Here, the CPLEX decoder denotes the decoder that exactly minimizes the energy $\mathcal{H}(C)$ defined in Eq.~\eqref{eq:energy} over error configurations consistent with the measured syndrome.
It should therefore be distinguished from full maximum-likelihood decoding over logical cosets.
However, for the strongly $Y$-biased case $p_x:p_y:p_z=1:10:1$ at $d=9$, even $N_\mathrm{SA}=100$ did not attain CPLEX-level accuracy.
This indicates that further improvements are needed to achieve CPLEX-level accuracy in this strongly $Y$-biased case.

We examined the convergence of the logical error rate for the number of inverse temperatures $N_\beta$ we employ for our SA decoder.
In particular, our SA decoder improves upon the initial configuration obtained from our greedy-matching decoder.
Furthermore, we confirmed that the convergence is strongly influenced by methods of constructing a reference configuration, which are employed to prepare initial configurations.
Especially, we proposed the ``Greedy matching (different)'' method, which may produce a different reference configuration for every invocation, and observed that it leads to markedly faster convergence.

Subsequently, we compared the elapsed times of various decoding algorithms.
In our sequential implementation, the SA decoder is faster than the CPLEX decoder when the physical error rate $p$ is relatively high, whereas it is slower in the lower-$p$ regime.
Because the independent SA tasks can be parallelized in principle, the idealized parallel-run-time estimate suggests that the SA decoder could become competitive with the CPLEX decoder.
This estimate, however, assumes perfect parallel efficiency, with parallel overhead neglected.
An actual parallel implementation and a quantitative study of the parallel overhead remain future work.

We list other topics for future work below.
Although our analysis has assumed the code capacity noise model, extensions to phenomenological and circuit-level noise models are important directions for future work.

The decoding time may be reduced by improving the MCMC algorithm.
Our SA decoder consists of independent SA processes, making it well-suited for parallelized implementation on FPGA.
Alternatively, techniques to accelerate the MCMC method could be employed, such as the replica-exchange Monte Carlo method~\cite{hukushima} and cluster algorithms including the Wolff algorithm~\cite{Wolff_algorithm}.
Furthermore, the Metropolis algorithm can be parallelized by domain (area) decomposition, which is implementable on FPGA~\cite{parallel_FPGA}.

Our SA decoder allows an explicit trade-off between decoding accuracy and run-time through tunable parameters such as the number of inverse temperatures $N_\beta$ and the inverse-temperature schedule $\{\beta_i\}_{i=1}^{N_\beta}$. 
In the present simulations, we did not use an adaptive stopping criterion.
We view developing systematic parameter-selection strategies for different code distances $d$, physical error rates $p$, and noise biases $p_x:p_y:p_z$, together with adaptive stopping criteria, as an important and nontrivial direction for future work.

An advantage of MCMC-based decoders is their generality: They require only the stabilizer generators and a specification of logical operators, without an explicit decoding graph. 
This makes them broadly applicable, e.g., to the color code and to quantum low-density parity-check codes~\cite{quantum_LDPC_codes}.
For SA initialization, one may use reference configurations produced not only by our greedy-matching decoder but also by other fast decoders, such as union find~\cite{union-find_decoder}.

Several decoders have been proposed that explicitly account for correlations induced by $Y$ errors~\cite{fusion_blossom,fowler2013optimalcomplexitycorrectioncorrelated,Paler_2023}. 
A systematic comparison with these approaches will be an important direction for future work.

Finally, we plan to perform simulations under device-specific Pauli-noise biases $p_x:p_y:p_z$. 
Toward practical deployment, further speed improvements will be particularly important in the low-$p$ regime, where the measured sequential SA run-time is currently longer than the CPLEX run-time.

\begin{acknowledgments}
We would like to thank Yusaku Takeuchi, Yugo Takada, Mitsuki Katsuda, Hiroshi Ueda, Hideaki Hakoshima, Kosuke Mitarai, Jun Fujisaki, Hirotaka Oshima, Shintaro Sato, and Keisuke Fujii for their helpful discussions.
This work was supported by the MEXT Quantum Leap Flagship Program (MEXT Q-LEAP), Grants No.\ JPMXS0118067394 and No.\ JPMXS0120319794; JST Moonshot R\&D, Grant No.\ JPMJMS2061; the JST COI-NEXT project Quantum Software Research Hub (Grant No.\ JPMJPF2014); and the JST COI-NEXT project Center of Innovation for Sustainable Quantum AI (Grant No.\ JPMJPF2221).
\end{acknowledgments}

\section*{Data Availability}
There are no publicly available research data or software supporting this manuscript.
Requests for further information or data should be sent to the authors.

\appendix

\section{Decoder formulated as integer programming problem}
\label{sec:integer_programming}

In this Appendix, instead of \Eqref{eq:determine_P_from_E}, we formulate the decoding problem as finding the recovery configuration $C$, subject to the constraint $\partial{C}=S$:
\begin{equation}
\arg\min_{C}\left\{\mathcal{H}(C) \,\middle|\, \partial{C}=S \right\}
.
\label{eq:optimal_error_chain_integer_programming}
\end{equation}

The error configuration $C$ can be represented as the set of one-bit binary variables $\{ x_i, y_i, z_i \}_i$ with the condition $x_i + y_i + z_i \le 1$ where $i=1,\dots,N_\mathrm{q}$ indexes the qubits.
Namely, $x_i=1$ indicates that the qubit $i$ suffers an $X$ error,
while the variables $y_i$ and $z_i$ similarly represent a $Y$ error and a $Z$ error, respectively.
Using this binary representation, the syndrome-consistency condition $\partial{C}=S$ in \Eqref{eq:optimal_error_chain_integer_programming} is explicitly expressed as a system of equations.
That is, for each face $f$,
\begin{equation}
\begin{split}
\left(z_{\mathrm{left}(f)} \oplus y_{\mathrm{left}(f)} \right) \oplus
\left(z_{\mathrm{right}(f)} \oplus y_{\mathrm{right}(f)} \right)&\\
\,\oplus
\left(x_{\mathrm{up}(f)} \oplus y_{\mathrm{up}(f)} \right) \oplus
\left(x_{\mathrm{down}(f)} \oplus y_{\mathrm{down}(f)} \right)&
= s_f
,
\label{eq:syndrome_consistency_binary}
\end{split}
\end{equation}
where $s_f \in \{0,1\}$ is the syndrome measurement outcome determined in \Eqref{eq:G_f_comm_anticomm}, and the operation $\oplus$ denotes xor (addition modulo two)\footnote{%
To solve Eq.~\eqref{eq:optimal_error_chain_integer_programming} with CPLEX, we convert the parity (xor) constraints in Eq.~\eqref{eq:syndrome_consistency_binary} into an equivalent integer linear programming using auxiliary integer variables and linear constraints.
}.
For boundary faces, where the stabilizer has weight three, the missing-qubit term is omitted from the parity (xor) constraint.

Under the constraint [Eq.~\eqref{eq:syndrome_consistency_binary}], the goal is to find the solution $\{ x_i, y_i, z_i \}_i$ that minimizes the objective function given by \Eqref{eq:energy}.
Here, the variables $n_x$, $n_y$, and $n_z$ represent the number of $X$-, $Y$-, and $Z$-type errors, respectively, and are given by
\begin{equation*}
n_x = \sum_{i=1}^{N_\mathrm{q}} x_i,\quad
n_y = \sum_{i=1}^{N_\mathrm{q}} y_i,\quad
n_z = \sum_{i=1}^{N_\mathrm{q}} z_i
.
\end{equation*}
This solution yields a recovery configuration $C$.

This problem falls into the category of binary integer programming, known to be NP-hard.
The integer programming formulation exhibits exponential computational complexity with respect to both the code distance $d$ and the physical error rate $p$.
Thus, for large values of $d$ or $p$, it is prohibitive to find a solution within a reasonable amount of time.
To solve the integer linear programming problem, we use CPLEX version 22.1.0~\cite{cplex}; we refer to the resulting decoder as the \emph{CPLEX decoder}.

The use of integer programming solvers for decoding was originally proposed and demonstrated for the color code~\cite{color_code_integer}.
The CPLEX decoder exactly solves the minimum-energy problem in Eq.~\eqref{eq:optimal_error_chain_integer_programming}, whose objective function is the energy $\mathcal{H}(C)$ defined in Eq.~\eqref{eq:energy}, which includes the probabilities $p_x$, $p_y$, and $p_z$.
The minimizer of this problem is a MAP error configuration among those consistent with the measured syndrome.
Thus, this minimum-energy formulation corresponds to the dominant-term approximation described in Sec.~IV~A rather than to full maximum-likelihood decoding over logical cosets.

\makeatletter
\immediate\write\@auxout{\string\citation{preprintControl}}
\makeatother
\bibliographystyle{apsrev4-2}
\bibliography{XZZX_SA_decoder}

\end{document}